\documentclass[twocolumn,floats,pra,showpacs]{revtex4-1}

\usepackage{amsmath}
\usepackage{amsfonts}

\usepackage{tikz}

\newcommand {\be}{\begin{equation}}
\newcommand {\ee} {\end{equation}}
\newcommand {\bea}{\begin{eqnarray}}
\newcommand {\eea} {\end{eqnarray}}
\newcommand{\non}{\nonumber}

\newcommand{\bE}{{\bf E}}

\newcommand{\cj}{{\cal J}}

\newcommand{\kj}{{\kappa_\cj}}
\newcommand{\kjs}{{\kappa_\cj^\sharp}}
\newcommand{\kjp}{{\kappa_\cj'}}

\newcommand{\nj}{{\nu(\cj)}}
\newcommand{\cR}{{\cal R}}
\newcommand{\cw}{{\cal W}}

\begin{document}


\title{Triviality of the ground-state metastate in long-range Ising spin glasses\\
in one dimension}
\author{N. Read}
\affiliation{Department of Physics, Yale
University, P.O. Box 208120, New Haven, CT 06520-8120, USA}
\date{January 17, 2018}

\begin{abstract}
We consider the one-dimensional model of a spin glass with independent Gaussian-distributed random
interactions, that have mean zero and variance $1/|i-j|^{2\sigma}$, between the spins at sites $i$ and $j$
for all $i\neq j$. It is known that, for $\sigma>1$, there is no phase transition at any non-zero
temperature in this model. We prove rigorously that, for $\sigma>3/2$, any translation-covariant
Newman-Stein metastate for the ground
states (i.e.\ the frequencies with which distinct ground states are observed in finite size samples in
the limit of infinite size, for given disorder) is trivial and unique. In other words, for given disorder
and asymptotically at large sizes, the same ground state, or its global spin flip, is obtained (almost)
always.
The proof consists of two parts: one is a theorem (based on one by Newman and Stein for short-range
two-dimensional models), valid for all $\sigma>1$, that establishes triviality under a convergence
hypothesis on something similar to the energies of domain walls, and the other (based on older results
for the one-dimensional model) establishes that the hypothesis is true for $\sigma>3/2$. In addition,
we derive heuristic scaling arguments and rigorous exponent inequalities which tend to support the
validity of the hypothesis under broader conditions. The constructions of various metastates are extended
to all values $\sigma>1/2$. Triviality of the metastate in bond-diluted power-law models
for $\sigma>1$ is proved directly.
\end{abstract}
\maketitle
\section{Introduction}
\label{intro}

The problem of the equilibrium properties of spin glasses has persisted for more than forty years. While
the basic standard model of a realistic short-range spin glass remains the Edwards-Anderson (EA) model
\cite{ea}, other models have also been considered, with the hope that they turn out to be more tractable.
Among these is a version of the EA model with power-law long-range interactions. In one dimension, the
Hamiltonian of this model has the form
\be
H=-\sum_{i,j\in{\bf Z}:i<j}J_{ij}s_is_j,
\label{1dham}
\ee
where $s_i$ are Ising spins, $s_i=\pm 1$, indexed by the set of integers $i\in\bf Z$, and as in the EA
model the bonds $J_{ij}=J_{ji}$ for pairs (or in a graph-theoretic language, ``edges'') $i$, $j$ are
independent Gaussian random variables with mean zero while, unlike in the EA model, the variances are
\be
{\rm Var}\,J_{ij}=\left\{\begin{array}{ll}|i-j|^{-2\sigma}\quad&\hbox{($i\neq j$),}\\
                                              0\quad&\hbox{otherwise.}
                                              \end{array}\right.
\ee
Here we allowed the summations to be carried out over an infinite system, though of course some
boundary conditions must be used to handle the infinite sum in practice. A periodic version of $H$ on
a system of length $L$ ($L>0$ an odd integer) can be constructed as follows. Let $i$,
$j$ be members of $\{-(L-1)/2,-(L-1)/2+1,\ldots, (L-1)/2\}$, let
$r_{ij}=\min(|i-j|, L-|i-j|)$, and let $J_{ij}=J_{ij}$ have variance $1/r_{ij}^{2\sigma}$ for
$|i-j|\neq0$, so for fixed $i$ and $j$ ${\rm Var}\,J_{ij}$ is $L$-independent when $L$ is
sufficiently large.
In this paper, the infinite size Hamiltonian (\ref{1dham})
is meant unless otherwise stated. We emphasize the spin-flip
symmetry of these Hamiltonians, that is, each Hamiltonian is unchanged if $s_i$ is replaced by $-s_i$
for all $i$ (i.e.\ flipping all the spins). One dimensional models have the advantage that the geometry,
especially
of domains, is much simpler than in higher dimensions, while the power-law form of the variance of the
interactions allows phase transitions to occur for sufficiently long-range interactions. The power-law
form, at least in higher dimensions, is also of interest because it can arise in realistic metallic
spin glasses.

In the one-dimensional power-law model, it is known rigorously \cite{ks,vEvH} that, in the parameter region
$\sigma>1/2$, the thermodynamic limit exists for thermodynamic properties when the temperature $T$ is
positive. For $\sigma\leq 1/2$, a non-trivial thermodynamic limit can be obtained (for temperature
$T$ held fixed in the limit) only if the Hamiltonian is first rescaled by an $L$-dependent factor; we do
not consider those cases in this paper. The absence of a transition at non-zero temperature (i.e.\
uniqueness of the Gibbs state) was proved by Khanin \cite{khanin} for $\sigma>3/2$. The classic
theoretical study of the one-dimensional model is by Kotliar, Anderson, and Stein \cite{kas}, who in
particular suggested that a transition at positive temperature would occur for $1/2<\sigma<1$ and not
for $\sigma>1$. This was followed by a proof of the absence of a transition at $T>0$ for $\sigma>1$ by van
Enter and van Hemmen \cite{vEvH2}. The last reference was criticized somewhat in Ref.\ \cite{cove},
in which the statement proved is the same but for fixed-spin boundary conditions only; see also Ref.\
\cite{vE}. Some further rigorous results were obtained in Refs.\ \cite{fz,gns}. For further theoretical
arguments from a physical perspective, see Refs.\ \cite{fh,bmy}, and for a selection of more recent
results on relevant issues, see Refs.\ \cite{ky,moore1d,mg,awmk}.

In the 1990s the theory of short-range spin glasses (especially the EA model) was revolutionized by the
work of Newman and Stein \cite{ns_rev}, who introduced the concept of a metastate to handle the possibility
of chaotic size dependence \cite{ns92}. A metastate is a probability distribution on Gibbs states of
an infinite-size system, derived from a limit of finite-size systems. At zero temperature, a ground-state
metastate (a probability distribution on ground states, of which there can be many when the system is
strictly infinite) might be trivial (all the weight on a single ground state, or a single pair if there
is spin-flip symmetry) or non-trivial (the weight is dispersed over many ground states not related by
symmetry). If replica symmetry breaking \cite{par79,mpv_book} occurs at non-zero temperature, the
metastate will be non-trivial \cite{ns_rev,read14}, even at zero temperature. In the scaling-droplet
theory \cite{bm1,macm,fh} (the main alternative scenario to replica symmetry breaking as a description
of a low-temperature spin-glass phase), the underlying assumptions imply that the metastate is trivial
in the low-temperature phase \cite{ns_rev}. The nature of the metastate is thus a basic issue for the
understanding of spin glasses. Some recent works have addressed metastates numerically \cite{wy,billoire}.

There is an (in general, unproved) expectation that any metastate, including a
ground-state metastate, will be trivial if there is no transition at $T>0$.
For the two-dimensional EA model (with only nearest-neighbor bonds on the square lattice), for which it
is widely believed that no transition occurs at non-zero temperature, Newman and Stein \cite{ns2d}
proved results that go part of the way towards showing that a translation-covariant zero-temperature or
ground-state metastate is trivial; we refer to this work as NS. The results were extended \cite{adns}
to show that such a metastate is indeed trivial in the case of a {\em half-}plane, rather than the full
two-dimensional plane.

In the present paper, we take up this topic in the case of the one-dimensional power-law model. The
original goal of this work, which will not be fully realized, was to prove that the ground-state
metastate is trivial for the model when $\sigma>1$. (This involves first constructing metastates for
the long-range model.) This does not seem to follow directly from the uniqueness of the Gibbs state at
$T>0$ for $\sigma>1$ mentioned above. The references \cite{khanin,vEvH2,cove,vE} are more concerned with
an infinite size limit at fixed $T>0$, and it does not seem possible to draw a conclusion about a
subsequent $T\to0$ limit. Finding the
ground-state metastate itself seems to require the opposite order of limits, namely $T\to0$ in finite size.

In order to prove a rigorous result, we follow NS's argument rather closely, and we carry it through to
prove triviality of any translation-covariant ground-state metastate (and uniqueness as well) for
$\sigma>1$, but only under
a hypothesis that certain energies (which are something like the energy of a domain wall) converge to
finite values. The hypothesis can be proved when $\sigma>3/2$ using results of Khanin \cite{khanin},
and combining these results proves that a ground-state metastate is trivial and unique for $\sigma>3/2$.
This is probably not optimal, and we also discuss scaling arguments intended to support the hypothesis
heuristically for all $\sigma>1$. The approach definitely will not work when there is a transition at
$T>0$ (in which case domain wall energies diverge), as is expected when $\sigma<1$, and the results shed
no light on whether replica symmetry breaking or a non-trivial metastate occur in the low-temperature
phases in that case.

Some additional side results obtained along the way are worth highlighting here, as they may be of
independent interest. One of these is a bound on the exponent, now usually called $\theta$, of the
scaling-droplet theory for the model \cite{bm1,macm,fh,bmy}, which describes the scaling of the minimum
domain wall energy (we define it more precisely below). We prove that it obeys the inequality
\be
\theta\leq \max(1-\sigma,0)
\ee
in the one-dimensional power-law model. Within the scaling-droplet theory, this gives a simple argument
that there can be no transition at $T>0$ for $\sigma>1$ (the known result discussed above).

Another side result is that we ultimately extend the NS construction of a ground-state metastate for the
model to all values $\sigma>1/2$, rather than $\sigma>1$ with which we begin, to obtain what we
call the ``natural'' ground-state metastate, and then extend the construction of an excitation metastate
(used in the proofs), and the proof of the main Theorem, likewise. These metastates will be useful when
considering other problems for the model within the metastate framework in the future.

In addition to these results, we also analyze variant models with diluted bonds, in which the probability
that a bond $J_{ij}$ is non-zero is a power of $|i-j|$. In these models, the triviality and uniqueness
of the ground-state metastate can be proved directly for all $\sigma>1$, similarly to the short-range
one-dimensional models.

The plan of the remainder of the paper is as follows. Section \ref{prelim} discusses various preliminary
matters: the almost-sure convergence of a certain sum of random variables, which physically is the
energy change when a finite set of spins is reversed; the rigorous definitions of Gibbs states,
ground states, and ground-state metastates; the notions of domains, superdomains, and rung energies, and
the issue of convergence of the latter; and the definitions of an excitation metastate, which extends
the ground-state metastate to include excitations, and of transition values and flexibilities. Section
\ref{theorem} states and proves the main Theorem of this work, the triviality of the ground-state
metastate under the hypothesis of  convergence of the rung energies. Section \ref{finite} first proves
that the hypothesis holds for
$\sigma>3/2$, then turns to heuristic scaling arguments which have some bearing on the convergence
question, and then back to rigorous arguments which allow the extension of the definition of
the metastates and of the Theorem to $\sigma>1/2$. Finally, the bond-diluted models are introduced and
analyzed.
The Appendix contains rigorous proofs of bounds on two scaling exponents that were discussed in Section
\ref{finite}.

\section{Preliminaries}
\label{prelim}

In this section, various preliminaries to the main Theorem are discussed: In Sec.\ \ref{prelim_basic},
we prove a useful basic lemma; in Sec.\ \ref{prelim_gibbs}, we explain Gibbs states, ground states, and
ground state metastates; in Sec.\ \ref{prelim_domain}, we define domain walls, microdomains, superdomains,
and rungs; and in Sec.\ \ref{prelim_excmeta} we introduce two technical tools: excitation metastates
and transition values.

\subsection{Basic observations}
\label{prelim_basic}

We let $S=(s_i)$ stand for the indexed collection (vector) of the values of all the $s_i$s, and
$\cj=(J_{ij})$ for the indexed collection (matrix) of all $J_{ij}$s. We also write $S(A)=(s_i:i\in A)$
for a subset $A\subseteq {\bf Z}$; $A^c={\bf Z}\backslash A$ is the complement of $A$ in $\bf Z$.
Following tradition, subsets of the one-dimensional lattice $\bf Z$
will also be denoted $\Lambda$, and $\Lambda_L=\{-(L-1)/2,-(L-1)/2+1,\ldots,(L-1)/2\}$ for odd $L>0$ is an
interval centered at the origin. We sometimes view the Hamiltonian as a function $H=H(S)$ of $S=(s_i)$
for given $\cj$. The probability distribution (measure) on $\cj$ will be denoted $\nj$ and is the infinite
product measure of the Gaussians for each $J_{ij}$. Note that we sometimes use the physicists' term
``distribution'' for the measure (or ``law'') \cite{chung,breiman} on some space (not always a space of
real variables); often ``distribution'' will mean the probability density (not the cumulative probability
on a single real variable), but this should be clear from
the context. For now, we assume a joint distribution $\nu\rho_\cj$ on the bonds $\cj$ and spins $S$, which
has marginal distribution $\nu$ on bonds $\cj$ (i.e.\ when we sum $\nu\rho_\cj$ over all $S$), and
conditional distribution $\rho_\cj$ on $S$ given $\cj$, which of course can depend on $\cj$ (later, a
metastate will be used in place of $\rho_\cj$). Measure-theoretic terms \cite{chung,breiman} such as
``with probability one'', or equivalently ``for almost all'' or ``almost surely'', refer here to the
joint measure $\nu\rho_\cj$ unless otherwise stated. We write $\bf P$ for probability of an event, and
$\bE$ for expectation (mean) and ${\rm Var}$ for variance of a random variable.

We will make frequent use of Kolmogorov's Three Series Theorem (Ref.\ \cite{chung}, p.\ 125), which
states that if $X_n$ is a sequence of independent random variables, and we define $Y_n=X_n\Theta(A-|X_n|)$
(where $\Theta$ is the step function with $\Theta(0)=1$, and $A>0$ is a constant), then
$\sum_{n=1}^\infty X_n$ converges almost surely if
and only if all of the following three numerical series converge: 1) $\sum_n{\bf P}[|X_n|>A]$, 2)
$\sum_n\bE Y_n$, and 3) $\sum_n {\rm Var}\, Y_n$. Then we can obtain the following result.
Consider the sum
\be
\sum_j J_{ij}s_is_j
\label{locfield}
\ee
for $i$ fixed, with $\cj$ and $S$ drawn from the joint distribution $\nu\rho_\cj$. The sum is bounded by
\be
\left|\sum_j J_{ij}s_is_j\right|\leq \sum_j |J_{ij}|,
\label{sum|J|}
\ee
which is independent of the spins, and likewise the differences of partial sums obey
\be
\left|\sum_{j=M+1}^N J_{ij}s_is_j\right|\leq \sum_{j=M+1}^N |J_{ij}|.
\label{partsum|J|}
\ee
\newline
{\it Lemma 1:\/} For $\sigma>1$ and with probability one, the right-hand side of (\ref{sum|J|})
converges and the sum (\ref{locfield}) converges absolutely, as does the sum $\sum_{i\in A,j\in
B}J_{ij}s_is_j$ where $A$, $B$ are subsets of $\bf Z$, $A \cap B=\emptyset$, and at most one of $A$, $B$
is infinite.
\newline
{\it Proof:\/}
This follows from the Three Series Theorem. The right-hand side of (\ref{sum|J|}) is independent
of the spins, so the problem reduces to one involving only the distribution $\nu$, to which the Three
Series Theorem can be applied with $A\to\infty$. We note that the series of expectation values
behaves as
\be
\sum_{j=1}^N \frac{1}{j^\sigma}\sim N^{[1-\sigma]_+},
\ee
up to a constant factor, where in writing such series we show the rate of divergence at the upper limit,
using the notation
\be
[x]_+=\left\{\begin{array}{ll}x\hbox{ if $x\geq 0$,}\non\\
                              0 \hbox{ otherwise,}\end{array}
\right.
\ee
or $[x]_+=\max(0,x)$, and omit the constant factor.
Likewise the series of variances behaves as
\be
\sum_{j=1}^N\frac{1}{j^{2\sigma}}\sim N^{[1-2\sigma]_+}.
\ee
(The convergence of the latter series is also the
condition for the existence of the thermodynamic limit \cite{ks,vEvH,cg_book}, which thus exists when
$\sigma>1/2$.)
Hence the right-hand side of ineq.\ (\ref{sum|J|}) converges almost surely, which gives the almost-sure
absolute convergence of the sum (\ref{locfield}), and its almost-sure convergence follows from the
Cauchy criterion and ineq.\ (\ref{partsum|J|}) in the usual way. The final statement follows immediately.
QED

\subsection{Gibbs states, ground states, and metastates}
\label{prelim_gibbs}

The results of this paper concern ground states of an infinite system, and probability distributions on
such ground states, called metastates. Here we collect the basic statements about ground states and
metastates. We begin, in a little more generality than we need, with Gibbs (or DLR) states at arbitrary
temperature, in a system that could be infinite (see e.g.\ Ref.\ \cite{bovier_book}). A state is a
probability distribution $\Gamma(S)$ on spin configurations $S$. A Gibbs state is defined via its
conditional distributions when conditioned on
$S$ outside a finite region $\Lambda$: namely if the conditional probabilities on the set
of $S(\Lambda)$, given $S(\Lambda^c)$, are $\Gamma(S(\Lambda)|S(\Lambda^c))$, then their ratios obey
\be
\Gamma(S(\Lambda)|S(\Lambda^c))/\Gamma(S'(\Lambda)|S(\Lambda^c))=\exp{-\beta (H(S)-H(S'))}
\ee
where $S'$ is another configuration, and $S'(\Lambda^c)=S(\Lambda^c)$, that is they are the same outside
$\Lambda$ ($\beta=1/T$ is inverse temperature). From this rule the full distribution $\Gamma(S)$
is determined by using an increasing sequence of $\Lambda$, but not uniquely; this allows the possibility
of many distinct Gibbs states, which can be thought of as arising from different choices of a ``boundary
condition'' (in a very general sense) at infinity. In short-range models, the difference $H(S)-H(S')$ under
this condition reduces to a finite sum, but in a long-range model it does not. In the one-dimensional model
(\ref{1dham}), by Lemma 1 the infinite sum converges absolutely almost surely for $\sigma>1$, so in this
case there is no issue with the definition of a Gibbs state. For $\sigma\leq 1$, a method of regularizing
the sum would have to be specified, and it seems that the existence of Gibbs states must be considered
further in this case. Fortunately we do not need to deal with this case in this paper.

The definition of a Gibbs state continues to apply when $T=0$, with (for $S\neq S'$) the ratio of
conditional probabilities being zero or infinity, unless the Hamiltonian difference is exactly zero,
which cannot occur when the distribution $\nu$ is continuous. In particular, for continuous $\nu$
the definition is satisfied in the case of a ground state, defined as follows: a configuration $S$ is
a ground state if changing $S$ on any finite subset $\Lambda$ of the spins causes a non-negative change
in energy. Again, with probability one there is no difficulty with this criterion in the one-dimensional
model when $\sigma>1$, and we note that in infinite size there can be many ground states, not all related
to one another by flipping all the spins. We mention that ground states are ``pure''
Gibbs states, but the definition of this notion in the general case $T\geq 0$ will not be needed here.
Even at $T=0$, not all Gibbs states are pure; a combination of ground states ia also a Gibbs state.
We let $\alpha$, $\beta$, \ldots label ground state configurations, namely $S^{(\alpha)}$, $S^{(\beta)}$,
\ldots of the infinite system, and write $\alpha=\beta$ if $S^{(\alpha)}=S^{(\beta)}$.

To deal with the possible multiplicity of infinite-size Gibbs states, and to make contact with finite
systems, we would like to use an infinite-size or ``thermodynamic'' limit of a sequence of finite systems,
all constructed from the same sample $\cj$. Because of the possible occurrence of chaotic size dependence
\cite{ns92}, this may not be straightforward. Newman and Stein proposed the use of a construct they called
a {\em metastate} \cite{ns96b,ns97}. A ground-state metastate is a probability distribution on infinite-size
ground states for the given $\cj$.
The existence of a metastate can be shown using empirical averages of finite-size systems: there exists a
sequence of sizes $L_k$, say, that is not dependent on $\cj$, such that the frequency of the number of
times each spin configuration $S(\Lambda_W)$ (observed in a fixed-size ``window'' $\Lambda_W$) occurs as
part of the ground state in the sizes $L<L_k$, with $W<L$, tends to a limit $\kj[S(\Lambda_W)]$ (the
probability of $S(\Lambda_W)$ under $\kj$) as $k\to\infty$, for arbitrarily large $W$. Any $S$ that
is drawn (sampled) from the metastate $\kj$ is a ground state: $S=S^{(\alpha)}$ for some $\alpha$.
(More precisely, in each size $L$, our Hamiltonian has a pair of ground states.
We write $\overline\alpha$ for the ground state with all spins reversed, namely
$S^{(\overline{\alpha})}=(-s_i^{(\alpha)})=\overline{S^{(\alpha)}}$, say. When we construct the metastate,
there are two ground states in each size $L$, and we actually use frequencies for the occurrence of
these pairs, but finally give each of $S^{(\alpha)}(\Lambda_W)$, $S^{(\overline{\alpha})}(\Lambda_W)$
equal probability in the metastate $\kj$.) Such a metastate is not known to
be unique---it might depend on the sequence $L_k$. In addition to this NS construction of a
metastate, there is also the earlier Aizenman-Wehr (AW) construction \cite{aw}. Both have similar
properties, and can be shown to be equal under some conditions \cite{ns96b,ns97}. The metastates
we use below could be of either type.

We should mention technically that a non-trivial
ground-state metastate might be spread over an uncountable number of
ground states, all but at most countably many of which have zero probability individually; in general
it will be {\em sets} of ground states that have non-zero probability. In particular, for any window of
finite size $W$, only a finite number $\leq 2^{W^d}$ (in space of dimension $d$) of ground states can
be distinguished in the window. A ground-state metastate assigns some non-zero probability to each
of these configurations in $\Lambda_W$. For a larger window, each of these ground states may resolve into
distinct ones, and share its probability between them. If we neglect the possible countable set of ground
states that can have non-zero probability individually, the situation in the uncountable case
is similar to that of a continuous distribution on the real numbers $x$ in say $0\leq x\leq 1$, with
the probabilities for
the restriction to a window corresponding to the probabilities for the restriction of the decimal
expansion of $x$ to a finite number of places after the decimal point. The probability of drawing a
given ground state (or number $x$) exactly will be zero, but it is still possible to speak of the
probability that a ground state (or number) drawn has some particular property, as long
as the latter is ``measurable''. We will not discuss measurability questions; see NS \cite{ns2d}.

In this paper, not many properties of a ground state metastate will be required, and the only property
used frequently is translation covariance. We can assume that the periodic model is used for the
finite-size systems, and then there is translation invariance of the joint distribution of
bonds and ground states in a given finite size, under translation of both the bonds and the spins. In
the limit defining a metastate, this translation invariance is inherited by the joint distribution
$\nu\kj$. The ground-state metastate $\kj$ is then translation covariant, that is, it is a distribution
on $S$ that depends on the parameters $\cj$, and is invariant under a translation of both the
spins $s_i$ and the bonds $J_{ij}$. If we take two metastates $\kj$, $\kjp$ for the given $\cj$, we can
draw two ground states independently (one from each), using the product distribution $\kj\kjp$. The
joint distribution $\nu\kj\kjp$ on bonds and pairs of ground states is also translation invariant if
the two metastates $\kj$, $\kjp$ are both translation covariant. Though translation covariance need
not hold in metastates in general, it is used in the proofs of the main results in this paper.

The construction of a ground-state metastate from finite-size samples still works for all $\sigma>1/2$, in
the sense that one obtains a distribution on spin configurations $S$. As the meaning of ground states in
infinite size is less clear for $1/2<\sigma<1$, we cannot say at the moment that these configurations are
infinite-size ground states. What we do know from the NS construction of a metastate is that their
restriction to any finite window occurs with non-zero frequency among finite-size ground states.
When the latter is all that is in use, a ground-state metastate in the present sense should be acceptable.
Many arguments still go through under these conditions. In most of this paper, we will not use this
approach, and will use the above definition where $\sigma>1$; we return to the more general approach again
in Section \ref{exten}.

\subsection{Domains and superdomains}
\label{prelim_domain}

From now on, we
frequently consider a pair of ground states, and whenever a probabilistic argument is used, we suppose
that we have two ground-state metastates for the same disorder $\cj$, say $\kj$, $\kjp$ (we could take
$\kjp=\kj$), and we draw the two ground states from these, say $\alpha$ from $\kj$ and $\beta$ from
$\kjp$ (independently). Measure-theoretic terms like almost all, almost surely, or with probability one,
now refer to the product of the measures $\nu$, $\kj$, and $\kjp$ with which bonds $\cj$ and ground
states $\alpha$ and $\beta$ are drawn, respectively.

By definition, if $S$ is a ground state, and any finite number (say, those in a finite set $\Lambda$) of
the spins are reversed, the energy change
\be
\Delta E=2\sum_{i\in\Lambda,j\in\Lambda^c}J_{ij}s_is_j
\ee
should be non-negative. By Lemma 1, this sum converges absolutely (with probability one) if $\sigma>1$.
We may sometimes describe this operation as the creation of the domain $\Lambda$, which in this case is
finite.

For two ground states $\alpha$ and $\beta$, the spins $s_i$ agree for some $i$ and not for others.
This defines two domains, which we denote by
$A$ (the set of sites where the spins are the same) and $B$ (the set of sites where the spins are reversed):
\bea
A&=&\{i:s_i^{(\alpha)}s_i^{(\beta)}=+1\},\\
B&=&\{i:s_i^{(\alpha)}s_i^{(\beta)}=-1\}=A^c.
\eea
(Here and in some following notation we suppress mention of the evident dependence on the two ground states
$\alpha$, $\beta$, which are typically fixed during the discussion.)
The domains $A$ and $B$ will necessarily both be infinite unless $\beta=\alpha$ or $\overline\alpha$
(otherwise there is a contradiction with the fact that $\alpha$ and $\beta$ are both ground states), and
the energy difference between $\alpha$ and $\beta$ will then be a doubly-infinite sum. If $\sigma<2$,
such a sum will usually not be convergent. In general, set $A$ is not (path-) connected, or contiguous:
its sites cannot all be reached from one another by a sequence of steps---each of length one---between
members of $A$. We call a domain consisting of an
interval of sites on which the spins are reversed compared with some reference state a ``microdomain''.
Hence $A$ (and likewise $B=A^c$) can be decomposed into a disjoint union of microdomains; the
intervals making up $A$ and those of $B$ of course alternate. The ``domain wall'' $\cal W$ between sets
$A$ and $B$ is the collection of nearest-neighbor edges with one end in $A$ and one end in $B$,
\be
{\cal W}=\{(i,i+1):s_i^{(\alpha)}s_i^{(\beta)}s_{i+1}^{(\alpha)}s_{i+1}^{(\beta)}=-1\},
\ee
which in general will contain many edges. In the case when a domain is a microdomain, we call the two
edges that bound it ``microwalls''; thus each edge $(i,i+1)\in{\cal W}$ is a
microwall. Precisely stated, $\cal W$ consists of either: (i) zero (when $\beta=\alpha$ or
$\overline\alpha$); (ii) a finite odd number (when $s_i^{(\alpha)}=s_i^{(\beta)}$
for all sufficiently large negative $i$, but $s_i^{(\alpha)}=-s_i^{(\beta)}$ for all sufficiently large
positive $i$, or {\it vice versa}); or (iii) an infinite number of microwalls. Case (ii) and some
instances of case (iii)
will be eliminated under some conditions as we go on. Clearly, each microwall $(i,i+1)\in{\cal W}$ is
of one of two types: we say $(i,i+1)$ is of type I if $i\in A$ and $i+1\in B$, and of type II if $i\in B$
and $i+1\in A$.

The strategy of the following proof (similar to NS \cite{ns2d}) will be to draw two ground states, say
$(\alpha,\beta)$ from a product $\kj\kjp$ of translation-covariant ground-state metastates. We
argue that if $\sigma>1$ and one condition is satisfied, then this leads to a
contradiction unless $\beta= \alpha$ or $\overline{\alpha}$. The conclusion will then be
that the metastates $\kj=\kjp$ are each supported on a unique ground state pair
$(\alpha,\overline{\alpha})$, or else $(\alpha,\beta)$ do not satisfy the condition. The argument involves
a ``superdomain'' of $\beta$ in $\alpha$, which we
now define: A superdomain of $\beta$ in $\alpha$ is a configuration $S$ that is identical to $S^{(\alpha)}$
for $i$ outside a finite interval such as $[i_0,i_1]=\{i_0,i_0+1,\ldots,i_1\}$ for $i_0<i_1$, but identical
to $S^{(\beta)}$ for $i$ in $[i_0,i_1]$.

We now begin to formulate the definitions that appear in the main arguments, and state the
condition used; they involve the change in energy when a superdomain from $\beta$ is inserted in
$\alpha$. If we write $A(i_0,i_1)=A\cap[i_0,i_1]$ and similarly $B(i_0,i_1)$, then this change in energy is
(recall that for $\sigma>1$, all these sums are finite with probability one, as a consequence of
Lemma 1)
\bea
\Delta E_{\beta(i_0,i_1),\alpha}&=&2\sum_{\stackrel{\scriptstyle i\in B(i_0,i_1),}{j\in B(i_0,i_1)^c}}
J_{ij}s_i^{(\alpha)}s_j^{(\alpha)}\\
&=&2\sum_{\stackrel{\scriptstyle i\in B(i_0,i_1),}{j\in A}}
J_{ij}s_i^{(\alpha)}s_j^{(\alpha)}\non\\
&&{}+2\!\sum_{\stackrel{\scriptstyle i\in B(i_0,i_1),}{j\in B\backslash
B(i_0,i_1)}}
J_{ij}s_i^{(\alpha)}s_j^{(\alpha)}.
\label{DeltaE}
\eea
The corresponding change in energy for the superdomain of $\alpha$ inserted into $\beta$
in the same interval is obtained by interchanging $\alpha$ and $\beta$ everywhere, and so is
\bea
\Delta E_{\alpha(i_0,i_1),\beta}&=&
2\sum_{\stackrel{\scriptstyle i\in B(i_0,i_1),}{j\in A}}
J_{ij}s_i^{(\beta)}s_j^{(\beta)}\non\\
&&{}+2\!\sum_{\stackrel{\scriptstyle i\in B(i_0,i_1),}{j\in B\backslash
B(i_0,i_1)}}
J_{ij}s_i^{(\beta)}s_j^{(\beta)}\\
&=&
{}-2\sum_{\stackrel{\scriptstyle i\in B(i_0,i_1),}{j\in A}}
J_{ij}s_i^{(\alpha)}s_j^{(\alpha)}\non\\
&&{}+2\!\sum_{\stackrel{\scriptstyle i\in B(i_0,i_1),}{j\in B\backslash
B(i_0,i_1)}}
J_{ij}s_i^{(\alpha)}s_j^{(\alpha)}.
\eea
Comparing these two energy changes, we see that the second sum on the right-hand side, the $BB$ terms, is
the same, but the first sum (the $AB$ terms) has reversed sign. Because only a finite set of spins changed
in both cases, both energy changes $\Delta E_{\beta(i_0,i_1),\alpha}$, $\Delta E_{\alpha(i_0,i_1),\beta}$
must be non-negative. Consequently, the $BB$ sum on the right hand side of each expression must be
non-negative, and the other sum, which has opposite sign in the two cases, must in magnitude be less than
or equal to the $BB$ sum.

We can also construct superdomains by using $\overline\alpha$ or $\overline\beta$ in place of {\em either}
$\alpha$ or $\beta$. A superdomain of $\overline\alpha$ in $\alpha$ is a single microdomain, with energy
$\Delta E_{\overline{\alpha}(i_0,i_1),\alpha}$ as above (in this case, $A$ is empty and $B={\bf Z}$).
Using the corresponding expressions, we obtain the identity (which will not be used in the paper)
\bea
\lefteqn{\!\!\!\!\!\!\!\!\!\!\!\Delta E_{\beta(i_0,i_1),\alpha}+\Delta E_{\alpha(i_0,i_1),\beta}{}+
\Delta E_{\beta(i_0,i_1),\overline{\alpha}}
+\Delta E_{\alpha(i_0,i_1),\overline{\beta}} }\qquad\qquad\qquad&&\non\\ \qquad&=&\Delta
E_{\overline{\alpha}(i_0,i_1),\alpha}+
\Delta E_{\overline{\beta}(i_0,i_1),\beta}.
\label{ident}
\eea

Some terminology for these combinations of energy differences will be useful. The $BB$ part of the sum for
$\Delta E_{\beta(i_0,i_1),\alpha}$ in eq.\ (\ref{DeltaE}) (which is the same as in $\Delta
E_{\alpha(i_0,i_1),\beta}$) corresponds to the interaction across the two ``rungs'' that form part of
a superdomain wall in the two-dimensional short-range case studied by NS. We may analogously call the two
edges $(i_0-1,i_0)$ and $(i_1,i_1+1)$, that bound the superdomain, rungs, even though the geometric
picture that applied in the two-dimensional short-range case does not apply here. A generic rung is denoted
by $\cal R$. The $BB$ sum itself we
call the ``total rung energy'' (that is, for both rungs together) of the superdomain. While the total rung
energy is finite (for $\sigma>1$), nothing prevents it from increasing as the length of the superdomain
increases. It will be crucial to assume as a hypothesis that the total rung energies (for different
superdomains constructed from the same two ground states $\alpha$, $\beta$) in some sense approach the
sum of two finite energies, one for each rung, as the separation of the rungs goes to infinity. Because
these energies are random, this must be defined carefully.

For simplicity, from here until Section \ref{exten} we will usually consider only rungs that are members of
the domain wall $\cal W$, and further, for a superdomain of $\beta$ in $\alpha$ or {\it vice versa},
we require $i_0$, $i_1\in B$. Then the rungs of our superdomain are $\cR_0$ of type I at the
left, and $\cR_1$ of type II at the right.

To define the energy of a single rung $\cR_0=(i_0-1,i_0)$ of type I, an obvious
approach would be to consider a semi-infinite analog
of a superdomain of $\beta$ in $\alpha$, with ground state $\alpha$ for the spins to the left of $i_0$,
and $\beta$ for $i_0$ and spins to its right. The rung energy for this single rung is
\be
2\sum_{\stackrel{\scriptstyle i<i_0,i\in B,}{j\geq i_0, j\in B}}J_{ij}s_i^{(\alpha)}s_j^{(\alpha)}.
\label{rungsum}
\ee
The next question is whether this sum converges to a finite value. Using the ``only if'' part of the Three
Series Theorem, one finds that the sum cannot be {\em absolutely}
convergent for $\sigma<2$ unless $B$ is very sparse, which we will see does not occur. Therefore we should
regularize the sum with a cut-off $R$ and then we ask whether these partial sums tend to a (finite) limit
as $R\to\infty$.
One ``symmetric'' way to regularize would be to restrict $i$ and $j$ to $\Lambda_R+i_0$, an interval of
length $R$ centered at $i_0$. But, instead of this, it will be useful to consider an asymmetric
regularization,
as follows. For the type I rung, we use the limit of
\be
E_{\cR_0,R}=2\!\sum_{\stackrel{\scriptstyle i<i_0,i\in B,}{j\geq i_0, j\in B,j\leq
i_0+R}}J_{ij}s_i^{(\alpha)}s_j^{(\alpha)}
\ee
as the regulating parameter $R\to\infty$, and similarly for the type II rung at $(i_1,i_1+1)$
we use the limit of
\be
E_{\cR_1,R}=2\!\sum_{\stackrel{\scriptstyle i\leq i_1,i\in B,i\geq i_1-R}{j> i_1, j\in
B}}J_{ij}s_i^{(\alpha)}s_j^{(\alpha)}
\ee
as $R\to\infty$. For finite $R$ and $\sigma>1$, these expressions converge absolutely
almost surely by Lemma 1. Moreover, when $R=i_1-i_0$, $E_{\cR_0,R}+E_{\cR_1,R}$
is exactly the total rung energy for the superdomain on the interval $[i_0,i_1]$.

If a rung energy $E_{\cR,R}$ (of either of the forms $E_{\cR_0,R}$ and $E_{\cR_1,R}$) converges almost
surely to a limit (denoted $E_{\cR_0}$ and $E_{\cR_1}$), then we say simply that the rung energy
$E_\cR$ converges. We now point out that the difference of the rung energies for any two rungs (for given
$\cj$, $\alpha$, and $\beta$) of the same type, when regularized with both $R$s tending
to infinity together (for example, with the $R$s equal) consists of two singly-infinite sums like those
discussed in Lemma 1, and for $\sigma>1$ one of these is absolutely convergent,
while the other tends to zero as $R\to\infty$, both with probability one. Hence for $\sigma>1$ and for
each type of rung, with probability one, either all rung energies are convergent, or all are non-convergent.
(This remains true if we consider rungs at arbitrary positions, and not only members of $\cw$, with now
two ways of regulating their energy instead of two types of rung. This is the reason that considering
only rungs in $\cw$ will be sufficient when $\sigma>1$.) This is still for given $\alpha$ and
$\beta$; it is possible that for given $\cj$, rung energies of either type converge for some pairs
$\alpha$, $\beta$ and diverge for others. Finally, it will be convenient
to use the convention that the case in which $\cw$ is empty, so the total rung energy is vacuous
(because there are no microwalls in $\cw$ to choose as rungs), falls into the class of non-convergence.

Similar issues arise for microwalls, which correspond to the case $B={\bf Z}$: the energy
change of a microdomain is finite, but the sum defining the energy change of either of its
two microwalls may not converge. Because of the identity (\ref{ident}), the scaling of the energy change
of a domain of $\overline{\alpha}$ in $\alpha$ with its length will be similar to that of the total rung
energy of a superdomain of the same size. Likewise, an identity similar to eq.\ (\ref{ident}) can be easily
obtained for semi-infinite superdomains, and gives a relation between the (formal) sums defining the energy
changes of a single rung and a single domain wall, and so the scaling of the regulated versions will also
be similar. These relations might be useful in further study of, for example, the expectation values of
these energy changes, to which we return in Sec.\ \ref{scaling}.

\subsection{Excitation metastate and transition values}
\label{prelim_excmeta}

The proof of the main Theorem involves the use of an excitation metastate.
Analogously to a ground state metastate, an excitation metastate \cite{ns2d} is a probability measure
on infinite-size excited
states and their excitation energies, where excited states are produced by constraining the values
$S(\Lambda)$ of the spins in finite sets $\Lambda$, in all possible ways and for all finite $\Lambda$,
and the energy of the constrained state is a minimum (for $\Lambda=\emptyset$, the excited state becomes
the ground state). It is produced from the limit of the finite-size version, like the ground state
metastate. In each finite size $L$ one can obtain such
excited states by constraining the values of the spins in $\Lambda$ (which must fit inside the finite
size), finding the lowest energy spin configuration subject to that constraint, and comparing its energy
with the ground state energy to obtain the energy change $\Delta E^{\Lambda,S(\Lambda),L}$. Note that
$\Delta E^{\Lambda,S(\Lambda),L}$ does not depend on which of the two flip-related unconstrained ground
states is chosen, and if $\Lambda$ is empty there are two (ground) states with $\Delta
E^{\Lambda,S(\Lambda),L}=0$. The frequencies of these sets of states and energies
$(\alpha_\cj^{\Lambda,S(\Lambda),L},\Delta E_\cj^{\Lambda,S(\Lambda),L})$ for all choices
$(\Lambda,S(\Lambda))$ (denoted $\sharp$) for given $\cj$ have a limit along a
$\cj$-independent subsequence $L_k$; this limit is an excitation metastate $\kjs$, a probability
distribution on $(\alpha^\sharp,\Delta E^\sharp)$ (for all $\sharp$ at once) for given $\cj$.
(For the excitation energies, frequencies can be obtained by a standard procedure such as binning
the energy changes and using the frequencies for the bins, then finally refining the size of the bins
to obtain the distribution.)
We also claim that for any ground state metastate $\kj$, we can extend it to an
excitation metastate $\kjs$, such that the marginal distribution of $\kjs$ on ground states is $\kj$.
For this, we simply require that the sequence of sizes $L_k$ used to obtain $\kjs$ be a subsequence of
that used to obtain $\kj$. In addition to this construction of a NS excitation metastate, there is an
analogous construction for an AW excitation metastate. Translation-covariant such metastates (obtained
from the periodic model) will be used in the proofs of the main results.

Notice that in the limit, the number of spins that flip when going from the ground state $\alpha$ to
an excited state $\alpha^\sharp$ is not necessarily finite (indeed it cannot be finite for both ground
states $\alpha$ and $\overline{\alpha}$), but that nonetheless the excitation energy $\Delta E^\sharp$ will
not diverge, because in any finite size it is bounded by the sum $2\sum_{i\in
\Lambda,j\in\Lambda^c}J_{ij}s_i^{(\alpha)}s_j^{(\alpha)}$ which in the limit (conditioned on $\alpha$)
converges absolutely with probability one for $\sigma>1$, by Lemma 1. (We remark that NS \cite{ns2d}
mention the technical point of establishing ``tightness'' (see Ref.\
\cite{chung}, p.\ 94) for the family of finite-size distributions of $\Delta E^\sharp$, which follows
from the bound on the excitation energy just mentioned. We comment further on this and show that the
restriction to $\sigma>1$ is not needed at the end of this section and in Appendix \ref{app}.)
We emphasize that there is a distinction between the
excitations (so-named following NS \cite{ns2d}) considered here, which involve minimizing the energy change
subject to a constraint on some spins, and the energy change for reversing some spins in a ground state,
leaving all others fixed, with no minimization, as discussed in the previous Section. To avoid confusion,
we will not refer to the latter as excitations.

From an excitation metastate, one can obtain information about how a ground state changes as a finite
number of the bonds $\cj$ are changed \cite{ns2d}. Suppose that $D$ is a finite set of edges $(i,j)=(j,i)$
($i\neq j$) and that $\cj^D$ is a set of values for bonds on $D$ (in other words, a function
$D\to{\bf R}$). Let $\Lambda$ be the set of sites $i$ that are the endpoints of edges in $D$. Define
\bea
H_{\cj^D}(S(\Lambda))&=&-\sum_{(i,j)\in D}J_{ij}^Ds_is_j,\\
H_{\cj}(S(\Lambda))&=&-\sum_{(i,j)\in D}J_{ij}s_is_j,
\eea
which involve only spins at sites in $\Lambda$.
Then in finite size we consider the functions
\be
h_{S(\Lambda)}^{(L)}(\cj^D)=\Delta E_\cj^{\Lambda,S(\Lambda),L}+H_{\cj^D}(S(\Lambda))-H_\cj(S(\Lambda)).
\ee
Define $S^{*,(L)}_{\cj,\cj^D}(\Lambda)$ to be one of the two $S(\Lambda)$ that minimizes
$h_{S(\Lambda)}^{(L)} (\cj^D)$. Then the ground state for bonds $\cj[\cj^D]$, that is $\cj$ with bonds for
edges in $D$ replaced by $\cj^D$, in size $L$ is
\be
\alpha_{\cj[\cj^D]}^{(L)}=\alpha_\cj^{\Lambda,S^{*,(L)}_{\cj,\cj^D}(\Lambda),L}
\ee
and its spin flip. Similarly, in the limit of infinite size, when the bonds in a finite set $D$ are
changed, the ground state $\alpha_\cj$ changes to $\alpha_{\cj[\cj^D]}$ or its spin flip, where now
\be
h_{S(\Lambda)}(\cj^D)=\Delta E_\cj^{\Lambda,S(\Lambda)}+H_{\cj^D}(S(\Lambda))-H_\cj(S(\Lambda))
\label{h_eff}
\ee
[$H_{\cj^D}(S(\Lambda))$ and $H_\cj(S(\Lambda))$ are unchanged], and
\be
\alpha_{\cj[\cj^D]}=\alpha_\cj^{\Lambda,S^{*}_{\cj,\cj^D}(\Lambda)},
\ee
where $S^{*}_{\cj,\cj^D}(\Lambda)$ is one of the two $S(\Lambda)$ that minimizes $h_{S(\Lambda)}(\cj^D)$.

The simplest case of a change in the bonds is when $D$ is a single edge $(i,j)$, so
$\Lambda=\{i,j\}$ contains only two sites. In this case we are considering the change of $J_{ij}$ to $K'$
with all other $\cj$ fixed. As $K'$ varies in $\bf R$, the ground state changes just once, from $\alpha$
when $K'=J_{ij}$ to a state $\alpha^{\{i,j\},S(\{i,j\})}$, where $S(\{i,j\})$ is one of the two spin
configurations on $\{i,j\}$ for which $s_is_j$ is minus its value in $\alpha$. The value of $K'$ at which
this change takes place is called the transition value \cite{ns2d}, denoted $K_{ij}$. $K_{ij}$  depends
on $\alpha$ and on $\cj\backslash \cj^D$. However, $K_{ij}$ and the unordered pair $\alpha$,
$\alpha^{\{i,j\},S(\{i,j\})}$ do not depend on the value of $J_{ij}$ when the other couplings are fixed.
This implies that $J_{ij}$ and $K_{ij}$  are independent with respect to the measure $\nu\kjs$.
Finally, we define the ``flexibility'' of the edge $(i,j)$ by \cite{ns2d}
\bea
F_{ij}&\stackrel{\rm def}{=}&2|J_{ij}-K_{ij}|\label{flexdef}\\
&=&\Delta E_\cj^{\{i,j\},S(\{i,j\})}
\label{flexex}
\eea
[using the same $S(\{i,j\})$], which is thus the minimum energy needed to change $s_is_j$ from its value
in the ground state $\alpha$ when the collection of $(\alpha^\sharp,\Delta E_\cj^\sharp)$ for all
$\sharp$ [in particular, for $\Lambda=\emptyset$ and $(\Lambda,S(\Lambda))=(\{i,j\},S(\{i,j\}))$] has
been drawn from $\nu\kjs$; this change in state can be accomplished by changing $J_{ij}$ by at least
$F_{ij}/2$, with the appropriate sign. We remark again that the flexibility $F_{ij}$ is bounded above
by the energy change for reversing any {\em finite} domain in ground state $\alpha$ such that $s_is_j$
changes, and that for $\sigma>1$ such energies converge (by Lemma 1). Thus such upper bounds on $F_{ij}$
can be obtained from $\alpha$ alone, which can be drawn from the ground state metastate $\kj$,
the marginal of the excitation metastate $\kjs$. This point will be useful in the next section.

We want to point out here that the technical restriction to $\sigma>1$ is not in fact necessary for the
construction of an excitation metastate in this model. In Appendix \ref{app} we show that for $\sigma>1/2$,
the finite-size excitation energies $\Delta E^{\sharp,L}$ are of order one with probability one, uniformly
in $L$, and while not necessarily convergent as $L\to\infty$, the family of distributions of these energies
is tight.
Then the use of subsequence limits allows us to obtain, in the same way as before, an excitation
metastate $\kjs$ for $\sigma>1/2$, and this can extend the version of the ground state metastate for
$\sigma>1/2$ already mentioned. We will not return to this until Section \ref{exten}, and in the
meantime continue to use the construction above for $\sigma>1$.

\section{Theorem}
\label{theorem}

Now we can state the main Theorem of the paper, the proof of which occupies the remainder of this Section.
The overall set-up and the terms used were already defined in the previous section. Measure-theoretic
statements like almost all, almost surely, or with probability one again refer throughout to the product
of the translation-covariant measures $\nu$, $\kj$, and $\kjp$ with which bonds $\cj$ and ground states
$\alpha$ and $\beta$ (respectively) are chosen, unless otherwise specified.
\newline
{\it Theorem:\/} Suppose $\sigma>1$. Then, for ground states $(\alpha,\beta)$ drawn from the
translation-invariant product $\nu\kj\kjp$, convergent rung energies for rungs $\cR\in\cw$ of both types
almost surely do not occur; that is, either any translation-covariant ground-state metastate
$\kj$ is supported on the same ground state pair $\alpha$, $\overline{\alpha}$ (that is, the metastate
is trivial and unique, $\kj=\kjp$), or if ground states $(\alpha,\beta)$ (with $\beta\neq
\alpha$ or $\overline{\alpha}$) occur, then the rung energies of at least one type are non-convergent.

The proof parallels that of the result in NS \cite{ns2d}, and will follow from Propositions 2 and 3 below.
To aid the reader, the Propositions in this Section are numbered so that they correspond to the analogous
Propositions in NS \cite{ns2d}.

We begin with basic facts, following NS.
\newline
{\it Proposition 1:\/} With probability one, the domain wall $\cw$ defined by the ground states $\alpha$
and $\beta$ as in the Theorem has well-defined non-negative density; if the density is zero, $\cw$ is empty.
\newline
{\it Proof:\/} By construction, the measure $\nu\kj\kjp$ is translation invariant, and so is the resulting
measure on bonds $\cj$ and sets $\cw$. The bonds $\cj$ can be integrated over,
producing the marginal distribution for the wall $\cal W$, which is again translation invariant.
The empirical density of the wall in an interval of length $L$ can be defined as the fraction of
nearest-neighbor edges in the interval that are in $\cw$ (or as a sum of indicator functions for such
edges, normalized by the length $L$). By the ergodic theorem for
the translation group ${\bf Z}$ \cite{breiman}, the empirical density has a well-defined
translation-invariant limit as $L\to\infty$, which we may call the density (it can depend on $\cj$,
$\alpha$, $\beta$). If the density is zero, with probability one microwalls are not seen at all in any
finite subregion of the infinite system. QED.

Proposition 1 justifies (for translation-invariant measures $\nu\kj\kjp$) two claims made earlier: that
case (ii) in the description of $\cw$ does not occur, and that the subset $B$ cannot be sparse, unless it is
empty. Indeed the argument in the proof of Proposition 1 can also be applied to the subset $B$, showing
that it has non-zero density if it is non-empty.

Now we turn to rungs and their energies, as defined in Sec.\ \ref{prelim}. We make the standing
assumption that $\sigma>1$ for the remainder of this Section. We already saw that the total rung energy
of a superdomain must
be non-negative. We now suppose that, for each type, (all) the rung energies $E_\cR$ converge, and
consider the infimum of the rung energies of type I,
\be
{\textstyle{\inf_{\rm I} E_\cR}}\stackrel{\rm def}{=}\inf_{\stackrel{\scriptstyle\cR\in\cw,}{\cR\in{\rm
I}}}E_\cR
\ee
(where I stands for the edges in $\cw$ of type I) and similarly $\inf_{\rm II} E_\cR$ for type II.
Then, by a similar argument involving the ergodic theorem as in the proof of Proposition 1, for any
$\varepsilon>0$ and for each type of rung there must be a non-zero density of them with rung energy within
$\varepsilon$ of the infimum for that type.
For well-separated rungs (one of each type, with type I at the left), by convergence the sum of rung
energies approximates the total rung energy---see Section \ref{prelim} (and see the proof of Proposition 2
below for a similar argument in greater detail). As the latter is non-negative and $\varepsilon>0$ was
arbitrary, it follows that
\be
{\textstyle{\inf_{\rm I}E_\cR}}+{\textstyle{\inf_{\rm II}E_\cR}}\geq0.
\ee
Then there are only two possibilities for the rung energies when both types converge: either the infimum
for either type is zero, or the infimum for at least one type is positive. Propositions 2 and 3 deal with
each possibility:
\newline
{\it Proposition 2:\/} For ground states $\alpha$, $\beta$ as in the Theorem, there is zero probability
that the rung energies converge and $\inf_{\rm I} E_\cR=\inf_{\rm II}E_\cR=0$.
\newline
{\it Proposition 3:\/} For ground states $\alpha$, $\beta$ as in the Theorem, for each type of rung, there
is zero probability that the rung energies converge and the infimum of rung energies of that type is
positive.
\newline
The Theorem follows immediately from these two Propositions.

{\it Proof of Proposition 2:\/} Suppose that there is non-zero probability that the rung energies of
$\alpha$ and $\beta$ converge and that $\inf_{\rm I} E_\cR$ and $\inf_{\rm II}E_\cR$ are zero. Then
for any $\varepsilon>0$ and for each type of rung there exist rungs with
rung energy less than $\varepsilon$, and these have a non-zero density by the ergodic theorem (see the
proof of Proposition 1). Hence for a microwall $(i,i+1)\in\cw$, we can find such a rung of type I, closest
and to the left of $(i,i+1)$, and another of type II, closest and to the right of it; at most one of
these can be $(i,i+1)$ itself. The sum of the rung energies of these is at most $2\varepsilon$, but
the total rung energy of the corresponding superdomain with these rungs may be larger than that. However,
the hypothesis that all rung energies converge (for this $\alpha$ and $\beta$) implies that for any
rung $\cR$ (of either type) and $\varepsilon>0$, there is an $R_0(\varepsilon,\cR)$ such that
$|E_{\cR,R}-E_\cR|< \varepsilon$ for all $R>R_0(\varepsilon,\cR)$. If for given $\varepsilon$ we
choose $R_0(\varepsilon)$ such that there are rungs $\cR$ of both types with $E_\cR<\varepsilon$ and
$R_0(\varepsilon,\cR)<R_0(\varepsilon)$, then another application of the ergodic theorem implies that
there is a non-zero density of such rungs (of each type). Then we can replace each of the two rungs
chosen before with one of the latter rungs of the same type, distant by more than $R_0(\varepsilon)$
from $(i,i+1)$ [they still enclose $(i,i+1)$], and then the total rung energy is $<4\varepsilon$. [More
generally, rungs can be found so that the total rung energy is $<\inf_{\rm I}
E_\cR+\inf_{\rm II}E_\cR+4\varepsilon$.] Then the energy change for the superdomain in either $\alpha$ or
$\beta$ is $<8\varepsilon$. Hence, as $\varepsilon$ was arbitrary, the energy
change required to reverse the sign of $s_is_{i+1}$ is arbitrarily small
(and this is true for {\em any} microwall in $\cw$). By the remark after eq.\ (\ref{flexex}) in Sec.\
\ref{prelim_excmeta}, this contradicts Proposition 4 below, proving the Proposition.

Proposition 4 states the intuitively obvious fact that, when the distribution of bonds is continuous
as it is here, there is zero probability that the minimum energy required to reverse $s_is_j$ is
exactly zero. The formal statement involves the notion of transition value that was introduced in
the context of the excitation metastate in Section
\ref{prelim_excmeta}; the remaining statements and proofs in this Section use the extension
of the two ground state metastates $\kj$,
$\kjp$ to excitation metastates $\kjs$, $\kappa_\cj'^{\,\sharp}$, and measure-theoretic statements are
now usually with respect to the measure $\nu\kjs\kappa_\cj'^{\,\sharp}$. This makes no difference to the
statements of the results in Propositions 2 and 3.
\newline
{\it Proposition 4:\/} There is zero probability that, for $\cj$ and a collection of $(\alpha^\sharp,\Delta
E_\cj^\sharp)$ for all $\sharp$ sampled from $\nu\kjs$, any given coupling $J_{ij}$ is
exactly at its transition value, that is, that the flexibility $F_{ij}$ of $(i,j)$ is zero.
\newline
{\it Proof:\/} The proof follows exactly as in NS from the independence of $J_{ij}$ and the transition
value $K_{ij}$ (see Sec.\ \ref{prelim_excmeta}), together with the fact that $\nu(\cj)$ is continuous. QED.

Note that the proof of Proposition 2 implies that, under the hypotheses, the flexibility $F_{i,i+1}$ of
$(i,i+1)$ in either $\alpha$ or $\beta$ is zero, even though
$F_{ij}=\Delta E_\cj^{\{i,j\},S(\{i,j\})}$ generally has to be sampled from $\kjs$, and so $F_{ij}$
cannot usually be obtained from $\alpha$ (sampled from $\kj$) alone.

{\it Proof of Proposition 3:\/} Again following NS, we use the notion of
a ``super-satisfied'' bond $J_{ij}$. First, a bond $J_{ij}$ is called ``satisfied'' in ground state
$\alpha$ if $J_{ij}s_i^{(\alpha)}s_j^{(\alpha)}>0$ for $S^{(\alpha)}$. For given $\cj$, $J_{ij}$ will
be satisfied in {\em every} ground state if
\be
|J_{ij}|>\min(\sum_{k:k\neq i,j}|J_{ik}|,\sum_{k:k\neq i,j}|J_{jk}|);
\label{supersat}
\ee
such a bond is called super-satisfied. By Lemma 1, the sums on the right-hand side converge for $\sigma>1$.

Suppose that, with non-zero probability, rung energies of type I for $\alpha$ and $\beta$ converge,
and $E'=\inf_{\rm I} E_\cR>0$.
Then by Proposition 4 and the ergodic theorem, we can find a rung $\cR=(i_0-1,i_0)$ in $\cw$ of type I
and two edges $\cR_1=(i,i+1)$ and $\cR_2=(j-1,j)$ in $\cw$ (where the first is type II and the
second type I) where $\delta=E_\cR-E'$ is strictly smaller than the flexibilities of both edges $\cR_1$
and $\cR_2$ in both $\alpha$ and $\beta$,
and where $\cR$ lies strictly between $\cR_1$ and $\cR_2$, that is $i+1<i_0-1<j-2$. Then the interval
$[i+1,j-1]$ contains both $A$ and $B$ sites. We now suppose that the bond $J_{i+1,j-1}$ is super-satisfied.
This means that no matter what other bonds with endpoints different from both $i+1$, $j-1$ are changed,
the spin product $s_{i+1}s_{j-1}$ cannot change sign. If not already true, this can be accomplished by
changing $J_{i+1,j-1}$ from its initial value, to increase $J_{i+1,j-1}s_{i+1}s_{j-1}$ without changing
$S$ and move away from the transition value $K_{i+1,j-1}$ (for both $\alpha$ and $\beta$); this does not
change the ground states $\alpha$ and $\beta$ because it does not change $s_{i+1}s_{j-1}$. The change
also does not change the rung energy of $\cR$ because $i+1$, $j-1$ are both in $A$,
and can only increase the flexibilities of $\cR_1$ and $\cR_2$ in either $\alpha$ or $\beta$, because it
only reduces the set of possible excitations.

Now we change $J_{ij}$ so as to reduce $2J_{ij}s_is_j$, that is we move $2J_{ij}$ towards its transition
value $2K_{ij}$ by an amount $\varepsilon$ (to be specified in a moment) slightly greater than $\delta$.
Usually, reducing $2J_{ij}s_is_j$ might cause either of the ground states $\alpha$, $\beta$ to change, due
to the appearance of an odd number of microwalls inside the interval $[i,j]$. Here, however,
such a change cannot produce an odd number of microwalls inside $[i+1,j-1]$ because $J_{i+1,j-1}$ is
super-satisfied. Provided we choose $\varepsilon$ larger than $\delta$ but smaller than the
flexibilities of the edges $\cR_1$ and $\cR_2$ in both $\alpha$ and $\beta$, it also cannot create a
microwall at $\cR_1$ or $\cR_2$ in either $\alpha$ or $\beta$. Thus reducing $2J_{ij}s_is_j$ by
$\varepsilon$ does not change $\alpha$ or $\beta$, but it does reduce the rung energy $E_\cR$ to below
$E'$, while leaving the energy of rungs $\cR'$ not contained in $[i,j]$ unchanged, so $E_{\cR'}\geq E'$.
Because the support of the distribution $\nu$ is unbounded as well as continuous (it is a product of
Gaussians), this gives a set of events with non-zero total probability that violate the ergodic theorem,
as discussed before Proposition 2.  QED.

\section{Convergence of rung energies and extensions of results}
\label{finite}

In this section, we first present arguments for the almost-sure convergence of the rung energies under
restrictions on $\sigma$. We begin with rigorous results, and then turn to heuristic ones. Then we describe
the rigorous extension of the main Theorem (but not the convergence of rung energies) to all $\sigma>1/2$,
and finally the analysis of models with diluted bonds.

\subsection{Rigorous convergence results}
\label{finite_rig}

The first result is simple to prove: if $\sigma>2$, then the rung energy (\ref{rungsum}), or the energy of
a single microwall (the same sum with $B={\bf Z}$), converges absolutely almost surely. This
result follows from the Three Series Theorem, similarly to that in Lemma 1, because the series of means and
of variances of $|J_{ij}|$ diverge as $L^{[2-\sigma]_+}$ and $L^{[2-2\sigma]_+}$, respectively.
It follows from the main Theorem that for $\sigma>2$, the ground state metastate is trivial and unique:
there are unique ground states $\alpha$, $\overline{\alpha}$ that carry the full probability in any
metastate. The Theorem and this result also apply to any model that has short-range, but not
necessarily nearest-neighbor, interactions. In these cases, the best one can do by more elementary
arguments is show that the number of ground states is bounded by a (calculable) constant of order one,
because for any finite region the spins outside impose at most a finite number of distinct boundary
conditions on it.

A stronger result can be obtained from work by Khanin \cite{khanin}. Probabilities are now evaluated
using $\nu$ on the space of $\cj$. The statement is:
\newline
{\it Proposition 5:\/} For $\sigma>3/2$, the rung energies of either type for any two ground states
$\alpha$, $\beta$ converge (in the manner defined in Section \ref{prelim}) almost surely.
\newline
{\it Proof:\/} We do not give all details, because almost all the work was done by Khanin for the same model
\cite{khanin}. The difficulty of evaluating the rung energy (\ref{rungsum}) (say for type I) for ground
state spin configurations is avoided by proving statements about {\em all} configurations, at the cost
of the restriction $\sigma>3/2$. His Lemma 3 states, for the case $B={\bf Z}$ (i.e.\ for a single
microwall---we return to the general
case afterwards) and with $R$ fixed, that for sufficiently large $R$ the probability that, for some $R'>R$
and some spin configuration $S$, the difference $|E_{\cR,R}-E_{\cR,R'}|$ is larger than a constant times
$R^{-t/2}$ is less than $\exp [-R^{(1+\delta/2)/2}]$, where $t=\sigma-3/2-\delta>0$ and $\delta>0$. Then
the probability that $\sup_S |E_{\cR,R}-E_{\cR,R'}|>\varepsilon>0$ for some $R$ and $R'$, where $R'>R>k$
for any given $k>0$ (the supremum is over all $S$),
is bounded by the sum over $R>k$ of the preceding probabilities. The sum converges (because the integral
$\int_0^\infty e^{-x^p}dx=p^{-1}\Gamma(1/p)$ for $p>0$ does), and so goes to zero as $k\to\infty$. That is,
for any $\varepsilon>0$,
\be
\lim_{k\to\infty}{\bf P}[\hbox{for some $R'>R>k$,}\; \sup_S|E_{\cR,R}-E_{\cR,R'}|>\varepsilon]=0
\ee
It follows (as in Ref.\ \cite{chung}, p.\ 70) that the microwall energy almost surely converges in the
sup norm $\sup_S|\cdots|$, or in other words for any $S$.

This leaves only a couple of points to settle in order to complete the proof. First, in the text Khanin
proves
his result for {\em bounded} random variables $J_{ij}$, but at the end of his paper indicates that it
also holds for Gaussian. The key lemma used is his Lemma 2, and we state the result for Gaussian
randomness in order to show how the restriction $\sigma>3/2$ first enters. First, for any fixed
configuration $S$, the probability that the magnitude of a sum
\be
\sum_{i\in F,j\in G}J_{ij}s_is_j
\ee
where $F$ and $G$ are intervals of length $a$, and separated by $a$, is larger
than a constant $\ell>0$ is less than $e^{-\ell^2 a^{2\sigma-2}/2}$. (This holds because
the sum is a Gaussian random variable with variance smaller than $a^{2-2\sigma}$, using the Chernoff bound;
see Ref.\ \cite{blm}, p.\ 22.) Then the probability that it exceeds $\ell$ for some $S$
is bounded by $2^{2a}e^{-\ell^2 a^{2\sigma-2}/2}$, which goes to zero as $a\to\infty$ if $\ell>
2a^{3/2-\sigma}$, and this is essentially Khanin's Lemma 2. Khanin's Lemma 3 is proved using multiple
applications of his Lemma 2, and $\sigma>3/2$ is needed so that the bounds $\ell$ being used can be small.
It also uses a bound on an infinite sum (similar to that in Lemma 1 above), which for Gaussian disorder
can be replaced by another similar bound on the probability of exceeding the former bound.

Finally, our rung energy is a similar sum but only spins in the subset $B\subset {\bf Z}$ enter the sums.
This can be incorporated by summing as before but viewing $s_i$ as taking the value $0$ if $i\not\in B$.
Then we have a model similar to the Ising case, but involving spins that take three values, $s_i=\pm 1$
or $0$. One can check that this makes no difference to Khanin's estimates (the $2^{2a}$ in the preceding
paragraph must be replaced by $3^{2a}$). Then the result is that convergence holds in the sup norm over
all configurations of Ising spins $S$ and choices of subsets $B$, which proves our Proposition 5. QED.

Together, Proposition 5 and the Theorem imply that a translation-covariant ground-state metastate of
the power-law one-dimensional model is trivial and unique for $\sigma>3/2$.

\subsection{Scaling arguments}
\label{scaling}

We expect that the restriction to $\sigma>3/2$ in the statement of Proposition 5 is not optimal, and that
the almost-sure convergence of the rung energies of either type, and hence the triviality and uniqueness
of the ground-state metastate, hold for all $\sigma>1$, provided that one considers only {\em ground
states} $\alpha$, $\beta$ rather than all spin configurations as in Khanin's results (his results
are not expected to hold for $\sigma<3/2$). We do not have a
rigorous argument for this, but we will discuss here some heuristic scaling considerations, which include
a small digression into more general questions, before addressing again the convergence of the rung
energies.

First we address a question that has probably occurred to readers: it is known that there is no transition
at non-zero temperature for $\sigma>1$; why does that not imply at least that domain wall energies are
finite? The intuition behind the question is that in one dimension, when reversing a domain of spins
of arbitrarily large length cost only a uniformly-bounded (and finite) energy, then at any positive
temperature domain walls will proliferate and destroy the order that was present in the ground state
(the rigorous proof for the case of a ferromagnet without disorder is in Ref.\ \cite{ruelle}). The
question is whether the converse to this statement holds.

There are two points to make in answering this. First, in the spin glass case, order would be destroyed
at positive temperature if there is a finite density of domain walls that are available for excitation
and have finite energy cost; it does not have to be the case that all domain wall (or microwall or
rung) energies must be finite---the energy of a microwall or rung energy at a given position could grow
more rapidly than that of the minimum energy ones. But in fact, we saw in Section \ref{prelim} that, if
$\sigma>1$, the energies of microwalls, like the energies of rungs, of either type either all converge
or all fail to converge, so then this question becomes moot.

Second (and extending the first point), order will not be destroyed when the temperature is
sufficiently low if the energy cost for creating a domain grows with distance, so that at sufficiently
low temperature the two domain walls are bound together. Thus positive temperature destroys the order
if (at least some of the) domain walls are unbound (or ``deconfined'') at arbitrarily low non-zero
temperature. If the energy of a domain scales as $E_n$ for a domain of length $n$, then schematically
the probability for the domain to have length $n$ at temperature $T$
is $e^{-\beta E_n}/Z_n$, where (in this paragraph) $\beta=1/T$ and the partition function is
\be
Z_n=\sum_{n\geq 1}e^{-\beta E_n}.
\ee
(This is schematic because we ignore other walls, and assume we can treat the domain with one end fixed,
as if there were translation invariance.) Then (similarly to bound states in quantum mechanics, for which
$Z_n$ corresponds to the norm-square of the wavefunction), if $Z_n$ is finite then the walls are bound,
and if it is infinite then they are unbound. This partition function (like all partition functions) is
a generalized Dirichlet series, of the general form $\sum_{n=1}^\infty a_ne^{-s \lambda_n}$, where $s$
is a parameter ($s=\beta$ here) and $\lambda_n$ are strictly increasing real numbers that tend to $\infty$.
In our case $a_n=1$, so the series diverges when $\beta=0$, and in this case with $\lambda_n=E_n$ the
series converges for ${\rm Re}\, \beta$ larger than
\be
\limsup_{n\to\infty} \frac{\ln n}{E_n}
\ee
(Ref.\ \cite{hardy}, p.\ 8), and note that this allows for the $E_n$ to be random variables that do not tend
to any limit. Hence if $\liminf E_n/\ln n=0$, domain walls are unbound at any $T>0$. Note that this
statement includes the preceding point that only the lowest energy domains are important asymptotically,
as well as well-known behavior of non-random models. Thus the absence of a transition at
positive temperature does not rigorously imply even that the smallest of the domain energies is always
finite, but only that they diverge at most sub-logarithmically, if they diverge at all. Nonetheless, it
does provide heuristic motivation for our conjecture.

For the energy change of a single microwall at a fixed position, for
example
\be
\sum_{i<0,j\geq0}J_{ij}s_is_j,
\ee
where $S$ is a ground state $\alpha$, and {\it a fortiori} for the rung energy (\ref{rungsum}),
the difficulty in estimating it is that the spins $s_i$ depend on the bonds $J_{ij}$. However, while
the spins certainly depend on the full set of $J_{ij}$, they may not depend strongly on all the bonds in
the smaller subset that occur in these sums. In particular, convergence of the sums is determined by the
tail at large $|i-j|$, and the corresponding bonds are weak, so the spins may not depend on them strongly.
If the spins are independent of the $J_{ij}$s, at least in the tail, then applying the Three Series Theorem,
the series of variances again diverges as $L^{[2-2\sigma]_+}$, while the expectation values are zero, and
so the energy of a single microwall would converge almost surely if and only if $\sigma>1$. We expect that
it may well be the case that the spins are approximately independent of the bonds in the sum with $|i-j|$
large, even for $\sigma<1$.

We now employ some scaling arguments (mostly obtained in finite size) to support this conjecture, but
focus only on the expected value of
the microwall energy. This involves the sum of terms $\bE[J_{ij}s_is_j]$ in a ground state, that is
(because $\bE[J_{ij}]=\bE[s_is_j]=0$) the correlation or covariance $C_{ij}$ of the bond $J_{ij}$ with the
corresponding spin product $s_is_j$. The obvious bound, from either the Cauchy-Schwarz or the Jensen
inequality, is
\be
C_{ij}=\bE[J_{ij}s_is_j]\leq \left(\bE [J_{ij}^2]\right)^{1/2}=\frac{1}{|i-j|^\sigma}.
\ee
If $s_is_j$ were independent of $J_{ij}$, the correlation would of course be zero, and if they were
approximately independent we would expect a small
value for large $|i-j|$, probably smaller than the preceding bound (i.e., a more negative power of $|i-j|$).

As discussed in Section \ref{prelim}, if $J_{ij}$ is changed, there will be a change to another ground
state when $J_{ij}$ passes $K_{ij}$, the transition value for the ground state $\alpha$ (this is also
the transition value as $J_{ij}$ approaches it from the other side, in the other ground state). For use
in the following, we will
make the scaling assumption that the transition values $K_{ij}$ (which are random variables that depend
on the two ground states involved, and hence on the other bonds) have a distribution of width
$|i-j|^{\theta'}$, for a ``transition value exponent'' $\theta'$ that presumably will not depend on
$\alpha$. The use of the symbol $\theta'$ is intended to suggest an analogy with the stiffness exponent
$\theta$ which has been defined for spin glasses, as we discuss in a moment. We can rigorously bound
$\theta'$ using a finite size system; see Appendix \ref{app}. The result is
\be
\theta'\leq0
\ee
for all $\sigma>1/2$.

There is a simple argument that shows that $\bE[J_{ij}s_is_j]\geq 0$ in a (finite-size) ground state
\cite{cg_book}
(introduce a parameter $\lambda$ by $J_{ij}\to\lambda J_{ij}$ with remaining $\cj$ unchanged, and use
the positivity of the second derivative of the expected value of minus the free energy at non-zero
temperature with respect to $\lambda$, then integrate with respect to $\lambda$ from $0$ to $1$ to
obtain the desired result, and finally take $T\to0$). Clearly when $|J_{ij}|$ is large, $J_{ij}s_is_j$
will be positive, and the sign of $s_is_j$ changes at $J_{ij}=K_{ij}$; thus $s_is_j$ equals the sign
of $J_{ij}-K_{ij}$:
\be
s_is_j={\rm sgn}\, (J_{ij}-K_{ij}).
\label{sgn}
\ee
This shows that while (as we argued in Sec.\ \ref{prelim_excmeta}) $K_{ij}$ and the
unordered pair of two ground states involved are independent of $J_{ij}$ when the other bonds are fixed,
at the same time the value of $J_{ij}$ selects one of the two ground states, and so is correlated
with $s_is_j$ such that $\bE[J_{ij}s_is_j]\geq 0$. If the transition value were zero almost surely,
or had a very narrow distribution ($\theta'<-\sigma$), then $C_{ij}$ would be of order
$|i-j|^{-\sigma}$. In the converse case $\theta'>-\sigma$ (which is the one we expect to occur)
in which the transition
value is typically relatively large, most of the weight in the Gaussian distribution of $J_{ij}$ falls
on one side of the transition value, giving only a small correlation. If the distribution of $K_{ij}$
has non-zero density near $K_{ij}=0$, then the probability that $K_{ij}$ falls in an interval
of order $|i-j|^{-\sigma}$ centered at $0$ will be of order $|i-j|^{-\sigma-\theta'}$. In this case
the correlation will be of order
\be
C_{ij}\sim|i-j|^{-2\sigma-\theta'}.
\ee
We can write both cases using the notation introduced in Sec.\ \ref{prelim}, as
\be
C_{ij}\sim |i-j|^{-\sigma-[\sigma+\theta']_+}.
\ee
This can also be obtained from an easy calculation using eq.\ (\ref{sgn})
(as a check on the result, the bound above from Cauchy-Schwarz is obeyed).

Next we would like to estimate or further bound the exponent $\theta'$. In fact it is easier to
consider the scaling of the flexibility $F_{ij}$ defined in eq.\ (\ref{flexdef}), which was identified
in eq.\ (\ref{flexex}) as the minimum energy that must be added to change the spin product $s_is_j$ from
its value in the ground state $\alpha$, for the original value of $J_{ij}$. From eq.\ (\ref{flexdef}),
we see that the ``flexibility exponent'' $\theta''$ for the scaling of the width of the distribution
of $F_{ij}$ with $|i-j|$ is equal to the larger of $-\sigma$ and $\theta'$, that is
\be
\theta''=\max(\theta',-\sigma),
\ee
and so $\theta''\leq0$. Hence it is equal to $\theta''=[\sigma+\theta']_+-\sigma$, and so for the
correlation
\be
C_{ij}\sim |i-j|^{-2\sigma-\theta''}.
\ee
The appearance of $\theta''$ in both places reflects the fact that both involve the scaling of
$J_{ij}-K_{ij}$, given by $\theta''$ as above.

The scaling of the minimum energy change for an excitation
that reverses the sign of $s_is_j$ compared with a ground state $\alpha$ (thus introducing an odd number of
microwalls between $i$ and $j$) is a question very similar to ones raised in the scaling-droplet theory
of spin glasses. There the lowest energy of a single domain wall (but not necessarily a
single microwall, in our language) that can be made in a region of size $L$ is supposed to scale as
$L^\theta$ \cite{bm1,macm,fh}. In the scaling-droplet theory, the sign of $\theta$ governs whether there
is a transition
at $T>0$, by arguments similar to the domain-wall binding discussed just above, but neglecting the
borderline cases we discussed there; thus $\theta>0$ means a transition at some $T>0$, while $\theta<0$
means there is none. The definition of $\theta$ can be made precise by defining $\theta$
as the scaling of the standard deviation for the change (which could be of either sign) in the
ground state energy when the periodic boundary condition is changed to antiperiodic in a system of size
$L$; such a change necessarily produces a single domain wall. In the power-law one-dimensional model
discussed here, this exponent obeys the bound
\be
\theta\leq\max(0,1-\sigma)=[1-\sigma]_+,
\label{thetabound}
\ee
as we prove in Appendix \ref{app}. If one ignores the possibility of logarithmic dependence for $\theta=0$,
this implies that there can be no transition at $T>0$ for $\sigma>1$.

We believe it is highly plausible that, when $\theta\leq0$, our exponent $\theta''=\theta$. The reason
behind the fact $\theta''\leq 0$ is that we require an excitation with an odd number of
microwalls between $i$ and $j$, which could be a single domain wall between them, but we do
not specify
whether there is one outside this interval. We also recall that $\theta$ is defined as the scaling of
the cheapest
energy for a single domain wall. If $\theta$ is negative, a second wall outside the interval could move
off to infinity and disappear, as this would lower the energy. But, if $\theta$ is positive, a flexibility
of order one could be obtained by placing domain walls just on either side of, say $i$ (or $j$); the
energy would be of order one because the domain size is order one (for $\sigma>1$ this is Lemma 1, but
for $\sigma<1$ it requires an argument given in Sec.\ \ref{exten} below, or we can appeal to the bound
$\theta'\leq0$ mentioned above). Hence we expect that in fact
\be
\theta''=\min(0,\theta)=-[-\theta]_+.
\ee
(This itself implies $\theta'\leq0$, because $\sigma>0$). It will be simplest to express scaling
relations in terms of $\theta''$.

If we now use the correlation $C_{ij}$ to calculate the expectation of the energy of a single microwall
at a given position, we find that it diverges as $L^{[2-2\sigma-\theta'']_+}$. Because the width of the
distribution of the minimum of a set of random variables must grow more slowly than the expectation of
each (even when they are not independent), and $\theta$ was defined as the scaling exponent for the energy
of the cheapest single domain wall in a region of length $L$, it must satisfy
\be
\theta\leq [2-2\sigma-\theta'']_+.
\ee
We can now consider cases. In general, $\theta''\leq0$. If $\theta''<0$, then $\theta=\theta''<0$. If
instead $\theta''=0$, then from the last displayed inequality $\theta\leq0$ if $\sigma>1$, which is a
contradiction unless $\theta=\theta''=0$. Hence in either case,
\be
\theta=\theta''\leq0 \quad\hbox{for $\sigma>1$,}
\ee
which also follows from earlier inequalities.

The expected ground state energy per site
is minus the sum of correlations $\sum_{j:j\neq i}\bE[J_{ij}s_is_j]$, and should converge for $\sigma>1/2$.
(For $\sigma<1$ this statement requires the further justification that we give in Sec.\ \ref{exten} below.)
By scaling,
we find that the sum diverges as $L^{[1-2\sigma-\theta'']_+}$. Then the exponent must be zero, giving
\be
\theta''\geq 1-2\sigma.
\ee
Altogether we then have
\be
0\geq\theta''\geq\max(-\sigma,1-2\sigma).
\ee
For $\sigma<1$, $\theta''\geq
1-2\sigma$ is the stronger lower bound, and in particular
\be
\theta''=\theta' \quad\hbox{for $\sigma<1$.}
\ee
The bounds pinch together as $\sigma\to 1/2$, which is the boundary of the region in which the thermodynamic
limit exists for thermodynamic properties.

It is known that $\theta=-1$ in the short-range model \cite{bm1} and the same is believed for the
present model when $\sigma>2$ \cite{fh,bmy}. In these cases the energy of a single microwall converges
almost surely, as we saw in Sec.\ \ref{finite_rig}.
In the nearest-neighbor short-range case, the energies for a microwall on different edges are
independent, and the cheapest one will scale as $L^{-1}$. This is expected to hold in the other
models mentioned also. As $\sigma$ decreases below $2$, we expect that $\theta$ increases from $-1$, subject
to the bound (\ref{thetabound}). For $1<\sigma<2$, where we expect that the energy for a single microwall
at a given fixed position converges (for example for $3/2<\sigma<2$, see Proposition 5 above), it is
again important that $\theta$ is defined as the {\em lowest} energy for a domain wall, which involves
minimizing over positions in an interval of length $L$ (it may also consist of more than one microwall).
A value $\theta >-1$ can occur presumably because of correlations among the energies of microwalls
at different positions, due to the long-range interactions. When the expectation of the single microwall
energy diverges as a positive power, it does so with exponent $2-2\sigma-\theta''$ which is at least
$2-2\sigma$, and $>0$ if
$\sigma<1$. This shows that in this regime the cheapest domain wall energy, for which the exponent is
$\theta\leq 1-\sigma$ for $\sigma<1$, is definitely less than than the expectation of the energy for
a single microwall at a given position; minimizing the energy makes a difference. In fact, we expect
that $\theta''=\theta'=0$ in the region $\sigma<1$.

It has been argued that for $\sigma<2$, $\theta=1-\sigma$ is an exact result \cite{fh,bmy,moore1d},
rather than only a bound as we find for $\sigma<1$. The arguments known to us for this conjecture do not
seem entirely convincing. For $\sigma<1$, the conjecture agrees with the scaling of the standard
deviation for the microwall energy [or with the bound (\ref{thetabound})], as if the spin products $s_is_j$
in the sum were independent of the bonds. However, recent work \cite{awmk} has argued that
$\theta=1-\sigma$ does not hold for $1/2<\sigma<2/3$, and that $\theta=1/6$ (the value obtained from
replica symmetry breaking) may hold there instead (correspondingly, $\theta=d/6$ is suggested
for $d>6$ in a short-range model in dimension $d$). Obviously, if correct, this undercuts the conjecture,
and if $\theta=1-\sigma$ holds for $\sigma>2/3$ it would mean that $\theta$ is a discontinuous,
non-monotonic function of $\sigma$ in the region $1/2<\sigma<1$, which we believe is unlikely. Numerical
work in Refs.\ \cite{awmk,ky,mg} is compatible with $\theta<1-\sigma$
for $2/3<\sigma<1$ as well as for $\sigma<2/3$.

In the region of interest to us here, $1<\sigma<2$, it is credible that $\theta=1-\sigma$ could be
exact, because all indications (including the results of the present paper) are that this region has
rather simple behavior. Again, numerical results are consistent with this for $\sigma$ not much larger
than $1$ \cite{ky}. This would then imply that $\theta''=1-\sigma>-\sigma$, so $\theta''=\theta'=1-\sigma$,
and then the expectation of the energy of a single microwall, or of a rung energy, at a fixed position would
converge. More generally, while a negative $\theta''$ could spoil the convergence of the expectation of
the single microwall energy, a value $0\geq\theta''\geq2-2\sigma$ would allow it to converge.

We may finally return to the original issue, the almost-sure convergence of the rung (or of microwall)
energies. The preceding arguments have suggested the scaling behavior of various properties, though
unfortunately without providing sufficiently strong bounds to settle the question (even granting the
scaling assumptions). They suggest that the expected value of the single microwall energy converges
for $\sigma$ not too small, while if $\theta''$ is not too negative, the correlation of $J_{ij}$ with
$s_is_j$ is weak. If the correlation of other long bonds $J_{kl}$ with $s_is_j$ are also weak for
$|i-j|$, $|k-l|$ large (an issue we have not addressed), then the conditions for the Three Series
Theorem could be ``almost'' met, leading to almost-sure convergence
for $\sigma>1$. Heuristically, the absence of a transition at $T>0$ strongly suggests this, and the
later arguments in this section are not much better than that.

\subsection{Extension to $\sigma>1/2$}
\label{exten}

Now we must argue, as promised, that the energy for creating any given finite domain (of fixed size)
in a ground state
is almost surely finite, not infinite (spins are never ``locked'' by the bonds), and that the ground state
energy per spin is likewise finite, for {\em all} $\sigma>1/2$, if the ground state
used is drawn from a ground-state metastate. (In other situations, for $\sigma<1$ spins may be locked;
see Ref.\
\cite{gns}.) We do not show that these domain energies converge in the infinite-size limit in general,
only that their distribution has no weight at infinite energy. We then build on these results to show how
the constructions of the various metastates used in this paper can be extended to all $\sigma>1/2$, and
extend the main Theorem likewise.

We recall that Lemma 1 proved the
absolute almost-sure convergence of the energy change when any finite set of spins
is flipped, provided $\sigma>1$. The probability measure used here is $\nu\kj$, and $\alpha$
is a ground state drawn from this measure; here $\kj$ is used in the more general conditions
$\sigma>1/2$ discussed at the end of Section \ref{prelim_gibbs}. For statements in finite size, we use
simply $\nu$.
We should define {\em tightness} of a family of probability measures $\mu_q$ ($q\in Q$
indexes the family) of a real random variable $X$ (Ref.\ \cite{chung}, p.\ 94): the family
is tight if for any $\varepsilon>0$ there exists an interval $I\subset {\bf R}$ such that
\be
\inf_{q\in Q}\mu_q(I)>1-\varepsilon.
\ee
More generally, for a family of measures on a general space, the interval $I$ is replaced by a compact set.
(Some authors use the term ``uniformly tight'' for what we call tight.)
\newline
{\it Proposition 6:\/} For $\sigma>1/2$, with probability one (i) the ground state energy density is a
finite constant, independent of $\alpha$;
(ii) there exists an excitation metastate $\kjs$ that extends $\kj$, and for any such extension the
excitation energies $\Delta E^\sharp$ [including the flexibility $F_{ij}$ of any edge $(i,j)$] are
almost-surely finite; and (iii) the family of finite-size joint distributions of the energy changes of any
finite collection of fixed finite domains is tight.
\newline
{\it Proof:\/} We first consider the ground state energy in finite size $L$, $E_0(L)=
-\sum_{i<j}J_{ij}s_is_j$, where we leave the finite-size cut-off on the summations implicit.
$E_0(L)=-\frac{1}{2}\sum_i\sum_{j:j\neq i}J_{ij}s_is_j$, and the local terms $e_i(L)=\sum_{j:j\neq
i}J_{ij}s_is_j\geq 0$ (in fact, $>0$ with probability one) because $2e_i(L)$  is the energy to flip the
single spin $i$. Note that for $\sigma<1$ it is not clear that $e_i(L)$ has a limit as $L\to\infty$, even
for a fixed
ground state configuration. Then the ground state energy per spin in finite size, ${\cal E}_0(L)=E_0(L)/L$,
is also a sum of negative terms. The ground state energy per site in the limit of an infinite-size system
is $\lim_{L\to\infty}{\cal E}_0(L)={\cal E}_0$ (if the limit exists). We are not aware of direct bounds on
this quantity, but there are lower bounds on the expected free energy per site
$\overline{f}=\lim_{L\to\infty}\overline{F(T)}/L$ at $T>0$, where $\bar{}$ denotes disorder average
$\bE$ and, for finite $L$, $F=-T\ln Z$, where $Z$ is the partition function; these bounds are used in the
proof \cite{vEvH,cg_book} that $\overline{F(T)}/L$ has a limit. Moreover, $F(T)/L$ tends to a limit almost
surely (by a variance bound $\sim L^{[1-2\sigma]_+-1}$ similar to those in Appendix $\ref{app}$, in which
the $T\to0$ limit can be taken). Unfortunately, the bound on $\overline{F(T)}/L$ is not uniform in $T$ as
$T\to0$, so it cannot be used to obtain a bound on $\overline{{\cal E}_0}$. However, we can appeal to
thermodynamics, which implies that $df/dT=-s(T)$,
where $s$ is the entropy per site (and we leave the $L\to\infty$ limit implicit). Both $-f$ and
$-\overline{f(T)}$ are convex functions of $T$, which implies that $s$ and $\overline{s}$ are defined
as functions of $T$ for almost all $T$ (in Lebesgue measure on $T$), and that $\overline{s(T)}$ is an
increasing function of $T$ (if there are values of $T$ at which $\overline{s(T)}$ jumps we can define
the value of $\overline{s(T)}$ there so that it is increasing at
all $T$). Moreover, ${\cal E}(T)=f+Ts$ is the internal energy per site, and $d{\cal
E}(T)/dT=Td\overline{s}/dT\geq 0$ is the specific heat (the heat capacity per site). (Here we define
the derivative for all $T$ by allowing it to include $\delta$-functions at jumps of $\overline{s}$; this
is legitimate as it will be integrated, not used as a function.) The entropy per site obeys $0\leq s\leq
\ln 2$, and $\overline{{\cal E}(T)}$ is finite at $T>0$. Integrating
\be
\int_0^T \frac{d\overline{s(T')}}{dT'} dT'=\int_0^T \frac{1}{T'}\frac{d\overline{{\cal E}(T')}}{dT'}dT'
\ee
gives a finite number $\overline{s(T)}$, so $d\overline{{\cal E}(T)}/dT$ can also be integrated over
the same range, showing that $\overline{{\cal E}_0}=\lim_{T\to0}\overline{{\cal E}(T)}=\lim_{T\to0}
\overline{f(T)}>-\infty$. The existence of the various limits and the finiteness of the expectation shows
that ${\cal E}_0$ is integrable (with respect to $\nu\kj$), and hence that the limit ${\cal E}_0$ exists
and is finite almost surely, which is statement (i).

For the second part of the proof, we return to the expectation of the ground state energy in finite size,
$E_0(L)=-\frac{1}{2}\sum_i e_i(L)$, so $\overline{E_0(L)}/L=-\frac{1}{2}\overline{e_i(L)}$ for any
fixed $i$ (by translation invariance), and we know
that this quantity has a finite limit. It follows from this, first, that the family (as $L$ runs over all
positive values) of probability distributions of $e_i(L)$ (induced from $\nu$) for any given $i$ is tight:
no weight goes off to infinity as $L\to\infty$. If it were not tight, by the definition above that would
mean that for some
$\varepsilon>0$ and for any finite interval $I=[0,\Delta]$, there would be some $L$ for which
${\bf P}[e_i(L)>\Delta]>\varepsilon$, and so $\bE[e_i(L)\Theta(e_i(L)-\Delta)]>\varepsilon \Delta$, which
goes to infinity as $\Delta$ (and hence also $L$) tends to $\infty$, contradicting finiteness of the
limit. Similarly, the family of {\em joint} distributions of $e_i(L)$ for $i$ in a fixed
finite set $\Lambda$ containing $n=|\Lambda|$ sites (with $L$ sufficiently large, so that $\Lambda$ is
contained in the system) is tight also. For this, it is sufficient to show that for any $\varepsilon>0$
there is some cube $C_\Delta=[0,\Delta]^n$ in the space of $n$-component
vectors $e(L)=(e_i(L))$ for $i\in\Lambda$ such that, for all $L$, ${\bf P}[e(L)\not\in
C_\Delta]\leq \varepsilon$. As
\bea
{\bf P}\left[e(L)\not\in C_\Delta\right]&=&{\bf P}\left[ \bigcup_{i\in\Lambda}
\left\{e_i(L)>\Delta\right\}\right] \\
&\leq& \sum_{i\in\Lambda}{\bf P}\left[e_i(L)>\Delta\right],
\eea
we can use the tightness of the families of distributions for each $e_i(L)$ (these distributions
are the marginals of the present joint distribution), putting $\varepsilon/n$ in place of $\varepsilon$
in the definition of tightness and using the corresponding value $\Delta$ (independent of $i$ by
translation invariance), to show that ${\bf P}[e(L)\not\in C_\Delta]\leq \varepsilon$.
The energy change for reversing the spins in a fixed finite set $\Lambda$, starting from the
ground state $\alpha$, differs from $\sum_{i\in\Lambda} e_i(L)$ only by a fixed finite set of terms
$J_{ij}s_is_j$, and so we obtain tightness of the family of joint distributions of these energy changes,
which is statement (iii). As similar
energies upper-bound the {\em minimum} excitation energies $\Delta E^{\sharp,L}$ for an excitation
$\sharp=(\Lambda,S(\Lambda))$, we then find that the distributions of the latter are tight as well.
This allows the
construction of an excitation metastate $\kjs$ extending $\kj$ as in Section \ref{prelim}, which is
statement (ii). (The last result is also obtained in Appendix \ref{app} by a
different method.) QED

The usual NS ground-state metastate does not include information on the energy change for reversing the
spins in a given finite domain in a ground state $\alpha$ drawn from the metastate. In the cases of the
EA model or the long-range one-dimensional model with $\sigma>1$, these energies can simply be calculated
from a ground state (because in the latter models the energy for reversing a finite set
of spins is almost surely finite and convergent, respectively). For $1/2<\sigma<1$, there is a convergence
issue if we take the infinite-size ground state $\alpha$ and attempt to take the limit as the
regularization (truncation) of the sum for the domain energy is removed. But now, because the
distributions of these energies are known to be tight, we can obtain (similarly to the construction of
the excitation metastate \cite{ns2d}) a ``natural'' extended metastate $\kappa_\cj^{\natural}$ which
gives the joint distribution of all such energy changes as well as ground state configurations. (As usual,
there is also an AW version of this construction.) We will
write $\alpha^\natural$ for a ground state spin configuration $\alpha$ augmented by
the collection of energy changes for reversing each fixed finite set of spins. It is now legitimate to
call the configuration $\alpha$ a ground state in infinite size, as all its possible energy changes
(needed when verifying that it is truly a ground state) have definite (non-negative) values
in an $\alpha^\natural$ drawn from the natural metastate $\kappa_\cj^{\natural}$.
It is natural that such a construction is required in the power-law model for $\sigma<1$, because the
convergence of the energy changes depends on the configuration (and the bonds) far away from the origin,
and this information is obtained by sampling from a (NS or AW) metastate.
This construction can also be combined with the construction of a metastate that gives minimum
excitation energies for constrained configurations, $\kjs$, to obtain another metastate
$\kappa_\cj^{\natural\sharp}$, which has both $\kappa_\cj^\natural$ and $\kjs$ as marginal distributions.

All constructions of superdomains in Section \ref{prelim} can now be carried through, simply by using
$\kappa_\cj^\natural$ in place of $\kj$, because these constructions simply involve reversing finite sets
of spins in either $\alpha^\natural$ or $\beta^\natural$.
This makes most uses of Lemma 1 unnecessary, and we can formulate the main Theorem with $\sigma>1/2$ in
place of $\sigma>1$. One use of Lemma 1 that cannot be replaced in this way was
its use to show that, for each type, either all rung energies (for given $\alpha^\natural$ and
$\beta^\natural$) converge,
or none do. This involves differences of rung energies; these differences are not energy changes
for reversing a finite number of spins in a ground state, so the sums are not finite sums of $e_i$, and
hence the approaches here
and in Appendix \ref{app} do not apply. Thus when stating and proving the Theorem, we cannot use that
result. Accordingly, we here consider rungs at arbitrary positions, not only in $\cw$. While an edge in
$\cw$ can still be classified as type I or type II, a rung may be used as the left or right end of a
superdomain, and the asymmetric regularization of its energy is defined accordingly, in one of two
ways, giving type I and type II rung energies for each rung (so one may converge
and the other not, for example). We now have the following extension of the main Theorem, where the
probability measure is $\nu\kappa_\cj^\natural\kappa_\cj'^{\,\natural}$ ($\kappa_\cj^\natural$ and
$\kappa_\cj'^{\,\natural}$ are natural metastates, extending $\kj$, $\kjp$, all for the same disorder
$\cj$):
\newline
{\it Theorem (extended):\/} Suppose $\sigma>1/2$. For any pair of translation-covariant metastates
$\kappa_\cj^\natural$,
$\kappa_\cj'^{\,\natural}$, there is zero probability that augmented ground states $\alpha^\natural$,
$\beta^\natural$ drawn from them have any convergent rung energies of both types, unless $\beta=\alpha$ or
$\overline{\alpha}$; that is, either any such metastate
$\kappa_\cj^\natural$ is supported on the same ground state pair $\alpha$, $\overline{\alpha}$, where
the energy changes do not have to be the same (that is, the underlying ground-state metastate is
trivial and unique, so $\kj=\kjp$), or if there are augmented ground states $\alpha^\natural$,
$\beta^\natural$ with $\beta\neq \alpha$ or $\overline{\alpha}$ in the metastate pair
then all the rung energies of at least one type are non-convergent, and further if some of
those of the other type are convergent then their infimum must be zero.
\newline
{\it Proof:\/} The extended Theorem follows as before from versions of Propositions 1--4, which however
must also be extended. Proposition 1 is unchanged, but also there is the parallel result that, for each
type of rung energy, the set of rungs whose energies of that type converge has non-zero density, or else
is empty. Propositions 2 and 3 must be extended,
as follows ($\sigma>1/2$, and the augmented ground states $\alpha^\natural$
and $\beta^\natural$ are obtained as in the statement of the extended Theorem):
\newline
{\it Proposition 2 (extended):\/} There is zero probability that some rung energies of both types I and II
converge and that the infimums of both sets of convergent rung energies are zero.
\newline
{\it Proposition 3 (extended):\/} For each type of rung energy, there is zero probability that some
converge and that the infimum of those that converge is positive.
\newline
The proof of the extended Proposition 2, using the sets of rungs with converging energies of either type,
then involves only minor changes of wording from that given before. It involves two rungs, one with
energy of each type, and an edge $(i,i+1)\in\cw$ between them, and also uses the extended Proposition 4.
The remaining statements and proofs involve the use of the extensions to metastates
$\kappa_\cj^{\natural\sharp}$, $\kappa_\cj'^{\,\natural\sharp}$. The statement and proof of Proposition 4
are unchanged. The proof of the extended
Proposition 3 involves a rung $\cR=(i_0-1,i_0)$ with convergent rung energy of one type, and the infimum of
the set of rung energies of this type is assumed to be nonzero. In addition, there are edges
$\cR_1=(i,i+1)$ and $\cR_2=(j-1,j)$, both in $\cw$, of types II and I as before, and arranged as before.
Super-satisfied edges do not exist for $\sigma\leq 1$, because the sums in ineq.\ (\ref{supersat}) diverge
by the Three Series Theorem, and do so with probability one \cite{chung}. Instead, for each of
$\alpha^\natural$ and $\beta^\natural$, we consider the
four sites $i$, $i+1$, $j-1$, $j$ as a set $\Lambda$, and the six edges between these sites as the set
$D$ (see Sec.\ \ref{prelim_excmeta}). Then we examine excitations to various $S(\Lambda)$ using the
excitation metastates, and consider transitions among these configurations as the bonds $\cj^D$ between
these sites are varied, similarly to the discussion of transition value and flexibility for a pair
of sites. For each of $\alpha^\natural$ and $\beta^\natural$, these are determined by an effective
Hamiltonian $h_{S(\Lambda)}(\cj^D)$ [see eq.\ (\ref{h_eff})] for the four spins which, apart from an
unimportant $S(\Lambda)$-independent function of the original bonds $\cj$ (determined by the ground
state $\alpha$ or $\beta$), contains the generalized transition values that can be viewed as six
two-spin, and a single four-spin, interaction terms (with finite coefficients), in addition to the
terms containing $\cj^D$.  The generalized transition values, or values of the effective couplings,
are independent of the six bonds $\cj^D$ among the four spins, like the transition values
earlier. Then $J_{i+1,j-1}s_{i+1}s_{j-1}$ can be increased sufficiently so that reducing $J_{i,j}s_is_j$
by the requisite $\varepsilon$ (with the other four bonds held fixed)
does not change $s_{i+1}s_{j-1}$ from its value in $\alpha$ and $\beta$, as in the earlier proof of
Proposition 3. The remainder of the proof is unchanged: Because of the unbounded support of the
probability distribution of the bonds, this leads to a non-zero probability for an event that violates
the ergodic theorem. This completes the proof of the extended Theorem.

The extended Theorem leaves open the possibility that, for augmented ground states $\alpha^\natural$,
$\beta^\natural$, rung energies of one type never converge, while some of those of the other type converge
and have zero infimum. Symmetry between the two types (i.e.\ reflection symmetry of the model) suggests
that this should not occur.

The extension of the Theorem removes the restriction to $\sigma>1$, but it does not remove the issue
of the convergence of the rung energies (of each type). It is highly unlikely that the rung energies ever
converge when $\sigma<1$. Hence the extended Theorem is probably not a great advance over the original one.
As the original Theorem is much simpler to state and to prove, that is the one we have emphasized.

\subsection{Bond-diluted power-law models}
\label{dilute}

In this section, we briefly consider a family of variants on the power-law model in one dimension that
have also been studied in the literature \cite{lprtrl}. In these models, we can establish both the absence
of a transition at positive temperature and the triviality of the metastate for $\sigma>1$ with some ease.

In these models, not all bonds are nonzero; instead the bonds are ``diluted''.
For $\sigma>1/2$, the probability that $J_{ij}$ is nonzero is taken to be
\be
p(i,j)=p_1/|i-j|^{2\sigma},
\ee
($0\leq p_1\leq 1$) independently for all pairs $i\neq j$ ($|i-j|$ can be replaced by
$r_{ij}$ in the finite-size variants); the non-zero bonds are assigned values for $J_{ij}$
drawn independently from a distribution (say, Gaussian) with variance one (when conditioned on being
non-zero), independent of their length $|i-j|$. Thus the variance of the bonds is ${\rm Var}\,
J_{ij}=p_1/|i-j|^{2\sigma}$ as before. Let us refer to non-zero bonds simply as bonds (i.e.\ bonds that
are present). The advantage of such a model for numerical purposes \cite{lprtrl} is that the expected
number of bonds ending at
$i$ is convergent, and so finite as $L\to\infty$, for all $\sigma>1/2$, unlike the original model.
The phase diagram, and the exponents at the transition with $T_c>0$ to the spin glass state for
$\sigma<1$, are expected to be essentially the same as in the original power-law model.

In this model in infinite size, the probability that site $0$ has no bonds connecting it to other sites is
\be
\prod_{j:j\neq 0,j\in{\bf Z}}\left(1-\frac{p_1}{|j|^{2\sigma}}\right),
\ee
which if $p_1<1$ converges to a non-zero value (rather than diverging to zero) if and only if the sum
\be
\sum_{j=1}^\infty\frac{1}{|j|^{2\sigma}}
\ee
converges. Hence for $\sigma>1/2$ and $p_1<1$ there is non-zero probability for a given site to
be disconnected, and there will be a non-zero density of such sites. Similarly, when $p_1<1$, there
will be finite sets of sites with no connection to the remainder of the system. Consequently for
$\sigma>1/2$ and $p_1<1$, these models have degenerate ground states with extensive entropy, and it will
be more appropriate to think of using a Gibbs state at zero temperature, rather than individual ground
states.

Similarly, the probability that there are no bonds between two halves of the system, say $i\leq 0$ and
$j>0$, is nonzero for $p_1<1$ if and only if the sum
\be
p_1\sum_{i\leq0,j>0}\frac{1}{|i-j|^{2\sigma}}
\ee
converges. As before, this diverges as $L^{[2-2\sigma]_+}$, and so converges if and only
if $\sigma>1$. [The sum also gives the expected number of bonds crossing $(0,1)$.] Then for $p_1<1$
and $\sigma>1$, by the ergodic theorem there is a non-zero density of
``cutting edges'', that is pairs $(i,i+1)$ that are not crossed by any bonds; the system breaks into
infinitely many disjoint intervals, each of finite length, that are not coupled to one another. This
implies immediately that in
these cases there is no transition at $T\geq 0$; the Gibbs state is unique at all temperatures, including
$T=0$, so the zero-temperature metastate is trivial and unique.

For $p_1=1$ and $\sigma>1/2$, for a continuous (e.g.\ Gaussian) distribution of bonds there will be
unique ground states, up to overall spin flip, in finite size with probability one,
as in the earlier models. For $p_1=1$ and $\sigma>1$, the probability that no bonds of length at least
$2$ cross a given edge $(i,i+1)$ is still non-zero, and there is a non-zero density of such edges; we
call the corresponding bonds of length one ``links''. We call the finite intervals between adjacent
links ``blobs''. A blob is an interval that cannot be further decomposed into intervals coupled only
by length-one bonds; it is either a single site, or has length $\geq 2$, and each blob is coupled to
its two neighbors by links. The system then behaves somewhat like the one-dimensional short-range model,
with blobs in place of single sites. At zero temperature, the links are all satisfied, like all the bonds
in the short-range model. The ground state (up to spin flip) can be found by first finding that of each
blob, ignoring the links, and then stringing together the ground states of the blobs, satisfying the
links, to obtain the ground state. Consequently, in infinite size the ground state is unique up to
a global spin flip, and the ground-state metastate is unique and trivial. We also observe that
the blobs-and-links picture for $\sigma>1$ implies that a microwall or rung energy is almost surely a
finite sum, and hence convergent. This implies that there is no transition at
$T>0$ in these models for any $\sigma>1$ when $p_1=1$ (as well as when $p_1<1$). For $\sigma\leq 1$, the
question of the metastate remains open, as for the other models (however, see also Ref.\ \cite{wy}).

Further, the blobs-and-links picture implies (heuristically) that the domain wall exponent $\theta$
takes the value $\theta=-1$ for all $\sigma>1$ when $p_1=1$. Comparing two finite-size ground states
that differ by reversal of the boundary condition (see Appendix \ref{app}), a domain wall can be made
at little energy cost by finding the link that is weakest (in magnitude), and the links are independent.
This gives the result of the short-range one-dimensional model as an upper bound, $\theta\leq -1$.
A domain wall could instead be created within a single blob, but the energy of a single microwall
is almost surely finite, regardless of location, and the blobs are independent. Some blobs may be large,
so the minimum microwall energy within a single blob might scale as a negative power of its length,
but it does not seem possible to arrive at $\theta<-1$, and we expect that $\theta=-1$. As a transition
at $T>0$ seems to occur when $\sigma<1$ \cite{lprtrl2}, implying $\theta\geq 0$ there, this
also implies discontinuous behavior of $\theta$ at $\sigma=1$.

While it is satisfying that in these models the ground state metastate is trivial for $\sigma>1$,
the very simplicity of the analysis, and the close relation with the short-range model, suggests that
in this regime these models may be a bit too simple to replace the original power-law model, which
required a deeper analysis.
However, these results might also suggest alternative approaches to the original model.

\section{Conclusion}
\label{conclusion}

To conclude, we have proved that a translation-covariant ground-state metastate of the power-law
one-dimensional spin glass model with exponent $\sigma$ (and zero magnetic field) is trivial and unique
for all $\sigma>3/2$. That is, only the same single pair of ground states will be seen (in any finite
window) in asymptotically large systems, with probability one. The main part of the proof is the Theorem,
which holds for all $\sigma>1/2$ and was proved following an argument of NS
\cite{ns2d}, but involves the hypothesis that the rung energies
converge, which has been proved only for $\sigma>3/2$. However, we suspect that the latter can be proved
for all $\sigma>1$, the region in which it is known that there is no transition at $T>0$. We also obtained
scaling arguments for related quantities, including rigorous bounds on scaling exponents such as $\theta$
in the scaling-droplet theory, and provided constructions of metastates for the model of wider interest.

The approach used to prove triviality of the metastate cannot work when there is a transition to a
spin-glass phase at some $T>0$, because then domain wall (and presumably rung) energies will diverge. At
the moment, a non-trivial metastate certainly cannot be ruled out in the low-temperature region
in those cases.

{\it Note Added:\/} Lemma 1 can easily be proved without invoking the Three Series Theorem, as follows.
If the sum of positive terms $\sum_j |J_{ij}|$ in ineq.\ (\ref{sum|J|}) diverged with non-zero probability,
then its expectation value would diverge, but for $\sigma>1$ the latter is finite, and so the sum
converges almost surely. The rest of the proof proceeds as before. The author thanks M. Aizenman for
pointing this out. There is a similar argument for the first result in Sec.\ \ref{finite_rig}.

\acknowledgments

The author is grateful for email discussions with G. Parisi (who raised
the question of the bond-diluted models) and M.A. Moore. This work was supported by NSF grant No.\
DMR-1408916.

\begin{appendix}
\section{Bounds on exponents $\theta$ and $\theta'$}
\label{app}

Here we prove the upper bounds on the exponents $\theta$ and $\theta'$ defined in Section \ref{scaling},
and some related results. This is included for completeness; the basic results are due to Aizenman and
Fisher \cite{af} and Newman and Stein \cite{ns92}, who considered $\theta$ in the EA model, and the method
is presumably similar to theirs. Here we work in finite-size systems, with probability distribution $\nu$.
In a fairly general form, the basic result is the following:
\newline
{\it Lemma 2:\/} if
$F_+$ is the free energy of an Ising spin glass Hamiltonian, and $F_-$ is the free energy for a
Hamiltonian that is the same except that $J_{ij}$ has been replaced with $-J_{ij}$ for the edges $(i,j)$
in a set that we call the ``cut'', and if the bonds $J_{ij}$ for edges in (or ``crossing'') the cut are
independent Gaussians with zero mean, then the variance of $F_+-F_-$ obeys
\be
{\rm Var}\,(F_+-F_-)\leq 4 \sum_{(i,j)\in {\rm cut}}{\rm Var}\, J_{ij}.
\ee
The statement holds more generally, provided that the bonds are independent, with
the distribution of each one invariant under reversing the sign of the bond.
It also holds similarly for other types of disorder in an Ising Hamiltonian, such as random fields
(single-site terms), or interactions involving $p>2$ spins, provided similar conditions hold.

To obtain the $\theta$ exponent bound from Lemma 2, we use the periodic boundary condition model. Let
the cut be the set of pairs $i<j$ with
$i\leq 0$, $j>0$, and $j-i<L/2$. Then $F_+-F_-$ as defined here is the change in free energy due to
reversing the boundary condition, and the bound is
\bea
{\rm Var}\,(F_+-F_-)&\leq&
4\sum_{\stackrel{\scriptstyle-L/2<i\leq0,0<j<L/2}{j-i<L/2}}\frac{1}{(j-i)^{2\sigma}}\\
&\sim& L^{[2-2\sigma]_+}
\eea
for large $L$. If the left hand side scales as $L^{2\theta}$, then we obtain
\be
\theta\leq \max(1-\sigma,0)=[1-\sigma]_+.
\ee
In a similar set-up in $d$ dimensions, in which ${\rm Var}\,J_{ij}\sim 1/r_{ij}^{2d\sigma}$
($\sigma>1/2$), and the
boundary condition is reversed in one of the $d$ directions on the hypercube, we obtain $\theta\leq
\max[d(1-\sigma),(d-1)/2]$ similarly. (This reduces to the bound $\theta\leq (d-1)/2$, obtained by the
authors cited, in the EA model with $d\geq 1$ or at $\sigma>1$, and is stronger than the bound
$\theta\leq d/2$ suggested for the power-law model in Ref.\ \cite{fh}.) Note that when $\sigma>1$, the
variance bound is stronger than that on
the exponent, $\theta\leq0$, as it rules out any diverging behavior for the domain wall free energy
including behavior slower than any power law, such as logarithmic or sub-logarithmic growth with $L$.

{\it Proof of Lemma 2:} We will provide a longer sketch of the proof than did Ref.\ \cite{ns92},
which indicated the idea. First, if $f(\tau_1,\ldots, \tau_M)$ is a function of random variables $\tau_I$,
$I=1$, \ldots, $M$, then its variance is
\be
{\rm Var}\, f= \bE f^2 - (\bE f)^2.
\ee
If $\bE_T$, where $T$ is a subset of $\{1,\ldots, M\}$, is expectation over the variables $\tau_I$ for
$I\in T$ with the remaining variables fixed (i.e.\ conditional expectation),
so $\bE=\bE_{\{1,\ldots,M\}}$, then by adding and subtracting terms we have
\bea
{\rm Var}\, f &=& \bE (f-\bE_{\{1\}}f)^2\non\\
              &&{}+\bE (\bE_{\{1\}}f-\bE_{\{1,2\}}f)^2\non\\
              &&{}\ldots+\bE (\bE_{\{1,\ldots,M-1\}}f-\bE_{\{1,\ldots,M\}}f)^2
\eea
(a ``martingale decomposition''). Each term can be viewed as a variance with respect to one variable
$\tau_I$ of $f_I$, where $f_I$ is already averaged over the $\tau_J$ for $J<I$, and with $\tau_K$ for $K>I$
held fixed, and a final average over $\tau_K$ for $K>I$.

The key observation in the case that $f=F_+-F_-$ (in which the $\tau_I$ are the $J_{ij}$s) is that
$\bE_{\{(ij)\in{\rm cut}\}}f=0$, provided only that the
joint distribution of these bonds is invariant under inversion (in particular, for independent bonds, if
the marginal distribution of each one is symmetric). By enumerating the bonds beginning with those in the
cut, this observation reduces the expression to a sum over terms that correspond to the
bonds in the cut only.

The $I$th term ($I=1$, \ldots, $M$) in the sum is the variance over $\tau_I$ with the later $\tau_K$
held fixed. For each term we use a standard bound for the variance of a Lipschitz function $f$ of a
single random variable $\tau$:  if $f$ has Lipschitz constant ${\cal L}$,
that is
\be
|f(\tau)-f(\tau')|\leq {\cal L}|\tau-\tau'|
\label{Lip}
\ee
for all $\tau$, $\tau'$ (${\cal L}<\infty$), then by averaging the square of eq.\ (\ref{Lip}) over $\tau$
and $\tau'$ (independently), we obtain
\be
{\rm Var}\, f\leq {\cal L}^2 {\rm Var}\,\tau.
\ee

For the free energy of an Ising spin glass, the Lipschitz constant is ${\cal L}=1$ for each random bond,
independent of the values of the other bonds. This follows by integrating the bound
\be
\left|\frac{\partial \ln Z}{\partial J_{ij}}\right|=\beta|\langle s_is_j\rangle|\leq \beta
\ee
(where again $\beta=1/T$) and if $J_{ij}$ is unbounded this is the best possible value. For the
difference of free energies, the Lipschitz constant is doubled. The same bounds hold for the expectation
of the
free energy, or differences thereof, over some other random variables. Assembling these facts proves
Lemma 2.

We can also apply the method of proof of Lemma 2 to obtain a bound on the width of the distribution of
transition values (defined in Sec.\ \ref{prelim_excmeta}). We single out one bond $J_{ij}$ and omit it
from the Hamiltonian. We fix the
spins $s_i$ and $s_j$, and let $F_+$ be the free energy for $s_is_j=+1$, $F_-$ the free energy for
$s_is_j=-1$
(obviously the spins $s_i$ and $s_j$ should not summed over in calculating these free energies in finite
size). At $T=0$, the difference of the corresponding ground state energies is $2K_{ij}$, that is, twice
the transition value for $J_{ij}$. The expectation of $F_+-F_-$ over only the bonds ending at $i$ or $j$
(these in effect constitute the cut) is zero, and then we find that the bound on ${\rm Var}\,
(F_+-F_-)$ is $\sim L^{[1-2\sigma]_+}$ (the distance $|i-j|$ does not enter the bound), and so is
bounded when $\sigma>1/2$. This implies immediately that $\theta'\leq 0$. In addition, by applying any
one of the standard Gaussian bounds for the tail of the probability distribution of a Lipschitz function
of Gaussian variables \cite{blm}, we find that the distribution of $K_{ij}$ (and hence also of $F_{ij}$)
is tight---no weight goes off to infinity as $L\to\infty$.

More generally, we can constrain the values of the spins on a finite set $\Lambda$, as used in the
excitation metastate. We can apply the same argument to the difference of free energies for two fixed
distinct configurations $S(\Lambda)$ on $\Lambda$, and find a similar bound by a constant that depends
on $|\Lambda|$. Thus (using a method similar to that in the proof of Proposition 6) the family (indexed
by $L$) of joint distributions of the differences of all such free energies as $S(\Lambda)$ runs through
the $2^{|\Lambda|}$ configurations are tight. Passing to zero temperature and the thermodynamic limit, we
can draw a ground state $\alpha$ from $\kj$, and one of the $S(\Lambda)$ coincides with
$S^{(\alpha)}(\Lambda)$. Then, by carrying this out for all finite $\Lambda$ and using subsequence limits,
we obtain an excitation metastate for any $\sigma>1/2$.

Finally we should mention that while the upper bounds on the variance, and on the tail of the distribution,
of a Lipschitz function go back further in the probability literature (see Ref.\ \cite{blm} for
references), a classic reference in the case of statistical mechanics of disordered systems is Ref.\
\cite{wa}, which also gives lower bounds on the variance.

\end{appendix}



\begin{references}


\bibitem{ea} S.F. Edwards and P.W. Anderson, J. Phys. F {\bf 5}, 965 (1975).

\bibitem{ks}K.M. Khanin and Ya.G. Sinai, J. Stat. Phys. {\bf 20}, 573 (1979).

\bibitem{vEvH} A.C.D. van Enter and J.L. van Hemmen, J. Stat. Phys. {\bf 32}, 141 (1983).

\bibitem{khanin} K.M. Khanin, Theor. Math. Phys. {\bf 43}, 445 (1980).

\bibitem{kas} G. Kotliar, P.W. Anderson, and D.L. Stein, Phys. Rev. B {\bf 27}, 602 (1983).

\bibitem{vEvH2} A.C.D. van Enter and J.L. van Hemmen, J. Stat. Phys. {\bf 39}, 1 (1985).

\bibitem{cove} M. Campanino, E. Olivieri, and A.C.D. van Enter, Commun. Math. Phys. {\bf 108}, 241
(1987).

\bibitem{vE} A.C.D. van Enter, J. Phys. A: Math. Gen. {\bf 21}, 1781 (1988).

\bibitem{fz} J. Fr\"ohlich and B. Zegarlinski, Commun. Math. Phys. {\bf 110}, 121 (1987).

\bibitem{gns} A. Gandolfi, C.M. Newman, and D.L. Stein, Commun. Math. Phys. {\bf 157}, 371 (1993).

\bibitem{fh} D.S. Fisher and D.A. Huse, Phys. Rev. Lett. {\bf 56}, 1601 (1986);
Phys. Rev. B {\bf 38}, 386 (1988).

\bibitem{bmy} A.J. Bray, M.A. Moore, and A.P. Young, Phys. Rev. Lett. {\bf 56}, 2641 (1986).

\bibitem{ky} H.G. Katzgraber and A.P. Young, Phys. Rev. B {\bf 67}, 134410 (2003).

\bibitem{moore1d} M.A. Moore, Phys. Rev. B {\bf 82}, 014417 (2010).

\bibitem{mg} C. Monthus and T. Garel, Phys. Rev. B {\bf 88}, 134204 (2013).

\bibitem{awmk} T. Aspelmeier, W. Wang, M.A. Moore, and H.G. Katzgraber, Phys. Rev. E {\bf 94}, 022116
(2016).


\bibitem{ns_rev} C.M. Newman and D.L. Stein, J. Phys.: Condens. Matter {\bf 15}, R1319 (2003).

\bibitem{ns92} C.M. Newman and D.L. Stein, Phys. Rev. B {\bf 46}, 973 (1992).

\bibitem{par79} G. Parisi, Phys. Rev. Lett. {\bf 43}, 1754 (1979); J. Phys. A: Math. Gen. {\bf 13},
L115 (1980); {\it ibid.}, {\bf 13}, 1101 (1980); {\it ibid.}, {\bf 13}, 1887 (1980).

\bibitem{mpv_book} M. Mezard, G. Parisi, and M.A. Virasoro, {\it Spin Glass Theory and Beyond}, Lecture
Notes in Physics, Vol.\ 9 (World Scientific, Singapore, 1987).

\bibitem{read14} N. Read, Phys. Rev E {\bf 90}, 032142 (2014).

\bibitem{bm1} A.J. Bray and M.A. Moore, J. Phys. C {\bf 17}, L463, L613 (1984); Phys. Rev. B {\bf 31}, 631
(1985); {\it Heidelberg Colloquium on Glassy Dynamics and Optimization}, Lecture Notes in Physics,
Vol.\ 275, eds.\ J.L. van Hemmen and I. Morgenstern (Springer-Verlag, Berlin, 1986), p. 121.

\bibitem{macm} W.L. McMillan, J. Phys. C {\bf 17}, 3179 (1984); Phys. Rev. B {\bf 30}, 476 (1984).

\bibitem{wy} M. Wittmann and A.P. Young,  J. Stat. Mech. Theor. Exp. {\bf 2016}, 013301 (2016).

\bibitem{billoire} A. Billoire, L.A. Fernandez, A. Maiorano, E. Marinari, V. Martin-Mayor, J. Moreno-Gordo,
G. Parisi, F. Ricci-Tersenghi, and J.J. Ruiz-Lorenzo, Phys. Rev. Lett. {\bf 119}, 037203 (2017).

\bibitem{ns2d} C.M. Newman and D.L. Stein, Commun. Math. Phys. {\bf 224}, 205 (2001).

\bibitem{adns} L.-P. Arguin, M. Damron, C.M. Newman, and D.L. Stein, Commun. Math. Phys. {\bf 300}, 641
(2010).

\bibitem{chung} K.L. Chung, {\it A First Course in Probability Theory}, 3rd Ed.\ (Academic, San Diego,
CA, 2001).

\bibitem{breiman} L. Breiman, {\it Probability} (Society for Industrial and Applied Mathematics,
Philadelphia, 1992).

\bibitem{cg_book} P. Contucci and C. Giardina, {\it Perspectives on Spin Glasses} (Cambridge University,
Cambridge, 2013).

\bibitem{bovier_book} A. Bovier, {\it Statistical Mechanics of Disordered Systems: A Mathematical
Perspective} (Cambridge University, Cambridge, 2006).

\bibitem{ns96b} C.M. Newman and D.L. Stein, Phys. Rev. Lett. {\bf 76}, 4821 (1996);  Phys. Rev. E {\bf 55},
 5194 (1997).

\bibitem{ns97} C.M. Newman and D.L. Stein, in {\it Mathematics of Spin Glasses and Neural Networks},
eds.\ A. Bovier and P. Picco (Birkhauser, Boston, 1997).

\bibitem{aw} M. Aizenman and J. Wehr, Commun. Math. Phys. {\bf 130}, 489 (1990).

\bibitem{blm} S. Boucheron, G. Lugosi, and P. Massart, {\it Concentration Inequalities: A Nonasymptotic
Theory of Independence} (Oxford University, Oxford, 2013).

\bibitem{ruelle} D. Ruelle, Commun. Math. Phys. {\bf 9}, 267 (1968).

\bibitem{hardy} G.H. Hardy and M. Riesz, {\it The General Theory of Dirichlet's Series} (Cambridge
University, Cambridge, 2015).

\bibitem{lprtrl} L. Leuzzi, G. Parisi, F. Ricci-Tersenghi, and J.J. Ruiz-Lorenzo, Phys. Rev. Lett. {\bf
101}, 107203 (2008).

\bibitem{lprtrl2} L. Leuzzi, G. Parisi, F. Ricci-Tersenghi, J.J. Ruiz-Lorenzo, Phys. Rev. B {\bf 91},
064202 (2015).

\bibitem{af} M. Aizenman and D.S. Fisher (unpublished; cited in Ref.\ \cite{ns92}).

\bibitem{wa} J. Wehr and M. Aizenman, J. Stat. Phys. {\bf 60}, 287 (1990).

\end{references}
\end{document}